\documentclass[a4paper]{article}

\usepackage[T1]{fontenc}

\usepackage{url}
\usepackage{hyperref}
\usepackage{authblk}
\usepackage{geometry}
\usepackage{graphicx}

\usepackage{xcolor}

\usepackage{amssymb,amsmath,amsthm}

\newtheorem{theorem}{Theorem}
\newtheorem{lemma}{Lemma}
\newtheorem{corollary}{Corollary}
\newtheorem{remark}{Remark}
\newtheorem{claim}{Claim}
\theoremstyle{definition}
\newtheorem{definition}{Definition}
\newtheorem{note}{Note}

\newcommand{\ie}{{\em i.e., }}
\newcommand{\eg}{{\em e.g., }}
\newcommand{\etal}{{\em et al.}}

\newcommand{\calL}{\mathcal{L}}
\newcommand{\calC}{\mathcal{C}}

\newcommand{\calG}{\mathcal{G}}

\newcommand{\peaceful}{\mathit{Peaceful}}

\newcommand{\ex}{\mathbf{E}}

\newcommand{\polylog}{\mathrm{polylog}}
\newcommand{\predef}{\stackrel{\mathrm{def}}{\equiv}}
\newcommand{\relmiddle}[1]{\mathrel{}\middle#1\mathrel{}} 
\newcommand{\longmid}{\relmiddle{|}} 
\newcommand{\xor}{\oplus}

\newcommand{\sch}{\mathbf{\Gamma}}

\newcommand{\outputs}{\pi_{\mathit{out}}} 

\newcommand{\call}{\mathcal{C}_{\mathrm{all}}}
\newcommand{\lone}{\mathcal{L}_{1}}
\newcommand{\lexist}{\mathcal{L}_{\ge 1}}
\newcommand{\lzero}{\calL_{0}}

\newcommand{\cpb}{\mathcal{C}_{\mathrm{PB}}}

\newcommand{\srl}{\mathcal{S}_{\mathrm{PL}}}
\newcommand{\cdl}{\mathcal{C}_{\mathrm{DL}}}
\newcommand{\cnz}{\mathcal{C}_{\mathrm{NZ}}}
\newcommand{\cmid}{\mathcal{C}_{\mathrm{mid}}}

\newcommand{\cdet}{\mathcal{C}_{\mathrm{det}}}

\usepackage[linesnumbered,ruled,vlined,noresetcount]{algorithm2e}
\DontPrintSemicolon
\newcommand{\mycommfont}[1]{{\footnotesize\ttfamily\textcolor{violet}{#1}}}
\SetCommentSty{mycommfont}
\SetArgSty{textnormal}
\SetKwProg{Fn}{function}{:}{end}
\SetKwProg{pFor}{for each}{ do in parallel}{end}
\SetKwInput{VarAgent}{Variables in Agents}
\SetKwInput{Macros}{Macros}
\SetKwInput{Input}{Input}
\SetKwInput{Notation}{Notation}

\newcommand{\colnum}{\xi}

\newcommand{\ppl}{P_{\mathit{PL}}}
\newcommand{\por}{P_{\mathit{OR}}}

\newcommand{\pcl}{P_{\mathit{CL}}}

\newcommand{\MT}{\mathtt{MoveToken}}

\newcommand{\determ}{\mathtt{DetermineMode}}
\newcommand{\elim}{\mathtt{EliminateLeaders}}
\newcommand{\create}{\mathtt{CreateLeader}}

\newcommand{\leader}{\mathtt{leader}}
\newcommand{\bull}{\mathtt{bullet}}
\newcommand{\shield}{\mathtt{shield}}

\newcommand{\signalb}{\mathtt{signal}_B}
\newcommand{\signalr}{\mathtt{signal}_R}

\newcommand{\ho}{\mathtt{clock}}

\newcommand{\hits}{\mathtt{hits}}
\newcommand{\id}{\mathtt{b}}

\newcommand{\distb}{\mathtt{dist}}
\newcommand{\cmax}{\kappa_{\mathrm{max}}}
\newcommand{\rmax}{\cmax}
\newcommand{\hmax}{\psi}

\newcommand{\token}{\mathtt{token}}
\newcommand{\tokenB}{\mathtt{token}_B}
\newcommand{\tokenW}{\mathtt{token}_W}
\newcommand{\mode}{\mathtt{mode}}

\newcommand{\iro}{\mathtt{color}}
\newcommand{\scol}{\mathtt{stampC}}
\newcommand{\sbin}{\mathtt{stampB}}

\newcommand{\cone}{\mathtt{c}_1}
\newcommand{\ctwo}{\mathtt{c}_2}

\newcommand{\dir}{\mathtt{dir}}
\newcommand{\strong}{\mathtt{strong}}

\newcommand{\Alg}{\mathtt{Alg}}
\newcommand{\Flip}{\mathtt{Flip}}

\newcommand{\tmp}{\mathtt{tmp}}

\newcommand{\target}{\mathit{Target}}
\newcommand{\valid}{\mathit{Valid}}

\newcommand{\Detect}{\mathit{Detect}}
\newcommand{\Construct}{\mathit{Construct}}

\newcommand{\last}{\mathtt{last}}

\newcommand{\distll}{d_{LL}}
\newcommand{\distrl}{d_{RL}}

\newcommand{\seqr}{\mathit{seq}_R}
\newcommand{\seql}{\mathit{seq}_L}

\newcommand{\prev}{\mathit{prev}}

\newcommand{\wl}{W_{\mathrm{LG}}}

\newcommand{\aspace}{\vspace{5pt}}
\newcommand{\varspace}{\hspace{10pt}}

\newcommand{\OmegaQ}{\Omega\text{?}}

\title{
A Nearly Time-Optimal Population Protocol for Self-Stabilizing Leader Election on Rings with Polylogarithmic States
\thanks{A preliminary extended abstract of this paper appears in the proceedings of PODC 2023.~\cite{podc}}}

\date{}

\author[1]{Daisuke Yokota}
\author[2]{Yuichi Sudo}
\author[3]{Fukuhito Ooshita}
\author[4]{Toshimitsu Masuzawa}

\affil[1]{The University of Osaka, Osaka, Japan}
\affil[2]{Hosei University, Tokyo, Japan}
\affil[3]{University of Hyogo, Hyogo, Japan}
\affil[4]{Notre Dame Seishin University, Okayama, Japan}

\begin{document}

\maketitle

\begin{abstract}
We propose a self-stabilizing leader election (SS-LE) protocol on ring networks in the population protocol model.
Given an integer $\psi$ satisfying $\log n \le \psi \le \log n+O(1)$, where $n$ is the population size, 
the proposed protocol reaches a safe configuration within $O(n^2 \log n)$ steps with high probability from any initial configuration,
and thereafter preserves the same unique leader forever.
Since no protocol solves SS-LE in $o(n^2)$ steps with high probability,
the convergence time is near-optimal, with only an $O(\log n)$ multiplicative gap. The proposed protocol uses only $\polylog(n)$ states.
Two state-of-the-art protocols are known for SS-LE on ring networks. The first protocol uses a polynomial number of states and solves SS-LE in $O(n^2)$ steps, whereas the second protocol requires super-exponential time but uses only a constant number of states. Our proposed protocol provides a useful middle ground between these two approaches.
\end{abstract}

\section{Introduction}
\label{sec:introduction}
This paper studies self-stabilizing leader election (SS-LE) in the population protocol model~\cite{AAD+06}.
In this section, we provide background and discuss related work in Section~\ref{sec:background},
present the contributions of this paper in Section~\ref{sec:contribution},
and explain how solving SS-LE in the population protocol model differs from solving it
in other standard models of distributed computing, such as the state-reading model and the message-passing model, in Section~\ref{sec:relation}.

Throughout this paper, $\log$ denotes the binary logarithm, i.e., $\log x = \log_2 x$,
and $[x,y]$ denotes the set of integers $\{i \in \mathbb{Z} \mid x \le i \le y\}$.

\subsection{Background and related work}
\label{sec:background}
In the population protocol model~\cite{AAD+06}, 
a network, called a \emph{population}, consists of a large number of finite-state automata,
called \emph{agents}.
Agents make \emph{interactions}
(i.e., pairwise communications) with each other
to update their states.
A population is modeled by a graph $G=(V,E)$,
where $V$ represents the set of agents,
and $E$ represents the set of possible interactions, that is, the pairs of agents that can interact.
Each pair of agents $(u,v)\in E$ interacts
infinitely often,
while each pair of agents $(u',v') \notin E$ never interacts.
At each time step,
one pair of agents chosen uniformly at random from all pairs in $E$ interacts.
Most studies in the population protocol model
make this assumption when evaluating the time complexities
of population protocols.
In the population protocol literature, many efforts have been devoted to devising protocols for complete graphs,
\ie populations in which every pair of agents interacts infinitely often,
while several studies
\cite{ARV25,AAF+08,AAD+06,BBB13,CP07,CC19,CG17,MNRS14,SOK+14,SMD+16,SOK+20det,SOK+18}
have investigated populations whose underlying graphs are not complete.

\emph{Self-stabilization} \cite{Dij74}
is a fault-tolerant property whereby,
even when any number of transient faults occur,
the network can autonomously recover from the faults.
Formally, self-stabilization is defined as follows:
(i) starting from an arbitrary configuration,
a network eventually reaches a \emph{safe configuration} (\emph{convergence}),
and
(ii) once a network reaches a safe configuration,
it maintains its specification forever (\emph{closure}).
Self-stabilization is of great importance in the PP model
because
this model typically represents 
a network consisting of a large number of inexpensive and unreliable nodes,
which requires strong fault tolerance.
Consequently,
many studies have been devoted to self-stabilizing population protocols \cite{AAF+08,BBB13,SIW12,CP07,CC19,FJ06,Izu15,SOK+14,SEI+21,SMD+16,SNY+12,SOK+20det,SOK+18,SOK+20polylog,YSM21}.
For example, Angluin, Aspnes, Fischer, and Jiang~\cite{AAF+08} proposed self-stabilizing protocols for a variety of problems, \ie leader election in rings, token circulation in rings with a pre-selected leader, 2-hop coloring in degree-bounded graphs, consistent global orientation in undirected rings, and spanning tree construction in regular graphs.
Sudo, Ooshita, Kakugawa, Masuzawa, Datta, and Larmore~\cite{SOK+18} gave a self-stabilizing 2-hop coloring protocol that uses much smaller memory space of agents. 
Sudo, Shibata, Nakamura, Kim, and Masuzawa~\cite{SSN+21} investigated the possibility of self-stabilizing protocols for leader election, ranking, degree recognition, and neighbor recognition on arbitrary graphs.

The self-stabilizing leader election (SS-LE) has received the most attention
among problems related to self-stabilization in the PP model.
This is because leader election is one of the most fundamental problems,
and several important protocols \cite{AAF+08,AAD+06,AAE08} assume a pre-selected unique leader.
In particular, Angluin, Aspnes, and Eisenstat~\cite{AAE08} show that 
all semi-linear predicates can be solved very quickly (\ie $\tilde{O}(n)$ steps with high probability) if we have a unique leader (although their protocol is not self-stabilizing).
Unfortunately, SS-LE is impossible to solve
without an additional assumption
even if we focus only on
complete graphs \cite{AAF+08,SIW12,SSN+21}. 
Many studies are devoted to overcoming this impossibility
by making an additional assumption, restricting topology, and/or slightly relaxing the requirements of SS-LE \cite{AAF+08,BBB13,BCC+26,SIW12, CP07, CC19,CC20,FJ06, Izu15,SOK+14,SEI+21,SMD+16,SNY+12,SOK+20det,SOK+18,SOK+20polylog,SSN+21,YSM21}.

Several studies~\cite{SIW12,BCC+26,BEG+25,GGS25,ABF+25,Sud26,SSN+21}
assume that every agent knows the exact number of agents.
Under this assumption, Cai, Izumi, and Wada~\cite{SIW12} gave an SS-LE
protocol for complete graphs, and Burman, Chen, Chen, Doty, Nowak,
Severson, and Xu~\cite{BCC+26} later improved its time complexity.
Subsequent studies have extensively investigated the time--space tradeoff
of SS-LE in this setting, including 
Berenbrink, Els{\"a}sser, G{\"o}tte, Hintze,
and Kaaser~\cite{BEG+25}, Gasieniec, Grodzicki, and
Stachowiak~\cite{GGS25}, Austin, Berenbrink, Friedetzky, G{\"o}tte, and
Hintze~\cite{ABF+25}, and Sudo~\cite{Sud26}.
In addition, Sudo \etal~\cite{SSN+21}
gave an SS-LE protocol for arbitrary graphs.

Several studies \cite{BBB13,CP07,FJ06} employ
\emph{oracles}, a kind of failure detector
to overcome the impossibility of SS-LE.
Fischer and Jiang \cite{FJ06}
introduced an oracle $\OmegaQ$ that eventually
tells all agents whether or not at least one agent satisfies a given condition,
e.g.~being a leader or having a special token.
They proposed two SS-LE protocols using $\OmegaQ$,
one for complete graphs and the other for rings.
After that,
Canepa and Potop-Butucaru \cite{CP07} proposed two SS-LE protocols that use $\OmegaQ$, \ie a deterministic protocol for trees
and a randomized protocol for arbitrary graphs.
Beauquier, Blanchard, and Burman \cite{BBB13}
presented a deterministic SS-LE protocol
for arbitrary graphs that uses two kinds of $\OmegaQ$.

Another class of studies \cite{AAF+08,CC19,CC20,FJ06,YSM21}
restricts the topology of a graph
to avoid the impossibility of SS-LE.
A class $\calG$ of graphs is called \emph{simple}
if there does not exist a graph in $\calG$ which
contains two disjoint subgraphs that are also in $\calG$.
Angluin \etal~\cite{AAF+08} proved that
there exists no SS-LE protocol
that works for all the graphs in any non-simple class.
If we focus on a simple class of graphs,
there may exist an SS-LE protocol for all graphs in the class.
As a typical example, the class of \emph{rings} is simple.
Angluin \etal~\cite{AAF+08} gave an SS-LE protocol that works
for all rings whose sizes are not multiples of a given integer $k$ (for example, rings of odd size).
They posed a question whether SS-LE is solvable or not for general rings (\ie rings of any size)
without any oracle,
while Fischer and Jiang \cite{FJ06} solved SS-LE for general rings using oracle $\OmegaQ$.
This question had been open for a decade until
Chen and Chen \cite{CC19} gave an SS-LE protocol for general rings.
These three protocols of \cite{AAF+08,CC19,FJ06} use only a constant number of states per agent.
The expected convergence times (\ie the expected numbers of steps required to elect a unique leader starting from any configuration)
of the protocols proposed by \cite{AAF+08,FJ06} are $\Theta(n^3)$, while the protocol of \cite{CC19} requires an exponentially long convergence time.
Here, the convergence time of the protocol of \cite{FJ06} is 
bounded by $\Theta(n^3)$ assuming that
the oracle immediately reports the absence of the leader
to each agent.
Yokota, Sudo, and Masuzawa \cite{YSM21} gave a time-optimal SS-LE protocol for rings: it elects a unique leader
within $\Theta(n^2)$ steps in expectation starting from any configuration
and uses $O(n)$ states,
given an upper bound $N$ on the population size $n$ such that $n \le N \le O(n)$.
Note that this knowledge is equivalent to an upper bound
$\psi$ such that $\log n \le \psi \le \log n+O(1)$, which we will assume in this paper.
Chen and Chen \cite{CC20} extended their protocol on rings so that it can work on arbitrary regular graphs.

Several studies \cite{Izu15,SOK+14,SEI+21,SMD+16,SNY+12,SOK+20det,SOK+18,SOK+20polylog} slightly relaxed the requirement of the original self-stabilization and gave \emph{loosely-stabilizing} leader election protocols. Specifically, the studies of this category allow a population to deviate from the specification of the problem (\ie a unique leader) after the population satisfies the specification for an extremely long time. 
This concept was introduced by \cite{SNY+12}.
The protocols of \cite{Izu15,SEI+21,SNY+12,SOK+20polylog}
work for complete graphs
and those of \cite{SOK+14,SMD+16,SOK+20det,SOK+18}
work for arbitrary graphs.

\begin{table}[t]
 \centering
 \caption{Self-stabilizing leader election on rings. Convergence time is shown in the expected number of steps. The assumption of~\cite{YSM21} was given as $n \le N \le O(n)$, which is equivalent to $\log n \le \psi \le \log n + O(1)$.}
 \label{tbl:results}
 \begin{tabular}[t]{c c c c}
  \hline
  &assumption/knowledge &convergence time\ \ & \#states\\
  \hline
 \cite{AAF+08} & $n$ is not multiple of a given $k$
      &$\Theta(n^3)$ & $O(1)$\\
 \cite{FJ06} & oracle $\OmegaQ$ & $\Theta(n^3)$ & $O(1)$\\
 \cite{CC19} & none & super-exponential & $O(1)$\\
 \cite{YSM21} & \ \ 
 $\log n \le \psi \le \log n + O(1)$
& $\Theta(n^2)$ & $O(n)$\\
 this work & \ \ 
 $\log n \le \psi \le \log n + O(1)$
& $O(n^2 \log n)$ & $\polylog(n)$\\
  \hline
 \end{tabular}
\end{table}

\subsection{Our contribution}
\label{sec:contribution}

We propose an SS-LE protocol $\ppl$ that works only for rings.
Specifically,
given an integer $\psi$ such that $\log n \le \psi \le \log n + O(1)$,
$\ppl$ elects a unique leader
in $O(n^2 \log n)$ steps both in expectation and with high probability
on any ring
and uses $\polylog (n)$ states.
The results for SS-LE on rings are summarized in Table~\ref{tbl:results}.
This work significantly reduces the number of agent states of \cite{YSM21} by slightly sacrificing the expected convergence time under the same assumption.
Our protocol is near time-optimal: it can be easily observed that no protocol solves SS-LE in $o(n^2)$ expected steps. 

The main technical contribution of this paper is a novel mechanism that
significantly reduces the number of states required
to detect the absence of a leader in the population compared to \cite{YSM21}.
The protocol of \cite{YSM21} requires $O(n)$ states to detect the absence of a leader, but our mechanism requires only $\polylog(n)$ states.
In addition, protocol $\ppl$ uses the same method as \cite{YSM21} to decrease the number of leaders to one.
As a result, protocol $\ppl$ requires only $O(n^2 \log n)$ expected steps,
while the existing three SS-LE protocols for rings \cite{AAF+08,CC19,FJ06} require $\Omega(n^3)$ expected steps.

Our protocol (and the protocol given in \cite{YSM21})
assumes that the ring is oriented or \emph{directed}.
\footnote{
The protocol with super-exponential
convergence time in \cite{CC19} is also designed for directed rings, but it works for undirected rings without sacrificing any complexity by using a known self-stabilizing ring-orientation protocol \cite{AAF+08} that uses only $O(1)$ states and requires a polynomial convergence time.
Unfortunately, we cannot use this ring-orientation protocol \cite{AAF+08} for our SS-LE protocol
because no suitable $O(n^2 \log n)$-time bound is known for the convergence time of this ring-orientation protocol.
}
However,
this does not lose generality because
we can remove this assumption: in Section~\ref{sec:orient}, 
we give a self-stabilizing ring orientation protocol that uses only a constant number of states
and requires $O(n^2 \log n)$ steps in expectation and with high probability.

\subsection{Relation with SS-LE protocols in the standard distributed computing model}
\label{sec:relation}
Generally, the model of population protocols is very different from the standard models of distributed computing such as the state-reading model and the message passing model. Unlike those models, in population protocols, (i) nodes are anonymous, (ii) nodes cannot distinguish their neighbors or even recognize their degree, and (iii) nodes update their states in pairs (they do not have the ability to observe the current states of all their neighbors when they update their states). These are limitations of population protocols. On the positive side, population protocols usually assume the random scheduler: at each time step, one pair of neighboring agents is chosen uniformly at random to interact. The random scheduler enables us to (loosely) synchronize the population and utilize randomization even with deterministic protocols. Our protocol fully exploits this assumption to achieve a fast convergence time with $O(\log \log n)$ bits per node.

When we focus on rings, the second weak point disappears because then every node knows that its degree is exactly two, thus the population protocols become closer to the standard models. To the best of our knowledge, every SS-LE protocol on standard models in the literature that has $O(\log \log n)$ bits and is designed for anonymous systems is slower than our protocol. The fastest one among such protocols is given by Awerbuch and Ostrovsky \cite{AO94} (Strictly speaking, some of those protocols have much smaller convergence time. However, we exclude them because they require strong knowledge such as the exact value of $m_n$, where $m_n$ is the smallest integer not dividing the network size $n$.) This protocol uses $O(\log^* n)$ bits and requires $O(n \log^2 n)$ asynchronous rounds to solve SS-LE. Even if we can simulate their protocol without additional cost in the model of population protocols, the convergence time would become $O(n^2 \log^3 n)$ steps in expectation, which is larger than our protocol. (Our protocol converges in $O(n^2 \log n)$ steps with $O(\log \log n)$ bits per node. Moreover, it requires $\Theta(n \log n)$ steps in expectation to let every node have an interaction at least once.)
Moreover, it requires careful discussion about the possibility to simulate protocols designed for ordinary models in population protocols.
In particular, when we use a small number of bits of agent memory such as $O(\log \log n)$ bits, 
there is no way in the literature to simulate the message passing model or state-reading model.

\section{Preliminaries}
A \emph{population} is a weakly connected digraph $G=(V,E)$,
where $V$ is a set of \emph{agents} and 
$E \subseteq V \times V$ is a set of arcs.
Define $n=|V|$. Each arc represents a possible \emph{interaction}
(or communication between two agents): If $(u,v) \in E$,
agents $u$ and $v$ can interact with each other,
where
$u$ serves as an \emph{initiator}
and $v$ serves as a \emph{responder}.
If $(u,v) \notin E$,
such an interaction never occurs. 
In this paper, we consider only a population represented
by a \emph{directed ring},
\ie we assume that $V=\{u_0, u_1, \dots, u_{n-1}\}$
and $E= \{(u_i,u_{i+1 \bmod n}) \mid i \in [0,n-1]\}$,
where $n \ge 3$.
Here, we use the indices of the agents only for simplicity of description,
and the agents cannot access their indices. The agents are \emph{anonymous},
\ie they do not have unique identifiers. 
We call $u_{i-1 \bmod n}$ and $u_{i+1 \bmod n}$ the \emph{left neighbor} and the \emph{right neighbor} of $u_i$, respectively.
We omit ``modulo by $n$''
(\ie $\bmod~n$)
in the indices of agents when no confusion occurs. 
We sometimes denote ``left to right'' direction by 
\emph{clockwise} direction, 
and ``right to left'' direction
by \emph{counter-clockwise direction}.
 
A \emph{protocol} $P=(Q,Y,T,\outputs)$ consists of 
a finite set $Q$ of states,
a finite set $Y$ of output symbols, 
transition function
$T: Q \times Q \to Q \times Q$,
and an output function $\outputs : Q \to Y$.
When an interaction between two agents occurs,
$T$ determines the next states of the two agents
based on their current states.
The \emph{output of an agent} is determined by $\outputs$:
the output of agent $v$ with state $q \in Q$ is $\outputs(q)$.
We assume that all agents share a common integer $\psi$ such that $\log n \le \psi \le \log n + O(1)$.
Thus, the components $Q$, $T$, $Y$, and $\outputs$ may depend on $\psi$. 
However, for simplicity,
we do not explicitly write protocol $P$
as parameterized by $\psi$, \eg 
$P(\psi) = (Q(\psi), T(\psi), Y(\psi), \outputs(\psi))$.

A \emph{configuration} is a mapping $C : V \to Q$ that specifies
the states of all the agents.
We denote 
the set of all configurations of protocol $P$ by $\call(P)$.
We simply denote it by $\call$ when
protocol $P$ is clear from the context.
We say that configuration $C$ changes to $C'$ by
an interaction $e = (u_i,u_{i+1})$,
denoted by $C \stackrel{e}{\to} C'$
if we have
$(C'(u_i),C'(u_{i+1}))=T(C(u_i),C(u_{i+1}))$
and $C'(v) = C(v)$
for all $v \in V \setminus \{u_i,u_{i+1}\}$.
We simply write $C \to C'$ if
there exists $e \in E$ such that $C \stackrel{e}{\to} C'$.
We say that a configuration $C'$ is reachable from $C$
if there exists a sequence of configurations
$C_0, C_1, \dots, C_k$ such that $C=C_0$, $C'=C_k$,
and $C_i \to C_{i+1}$ for all $i \in [0,k-1]$.
We also say that a set $\calC$ of configurations
is \emph{closed} if
no configuration outside $\calC$ is reachable from 
any configuration in $\calC$.

A scheduler determines which interaction occurs at each time
step (or just \emph{step}).
In this paper, we consider a \emph{uniformly random scheduler}
$\sch=\Gamma_0, \Gamma_1,\dots$, where each $\Gamma_t \in E$ is an independent random variable
such that $\Pr(\Gamma_t = (u_i,u_{i+1})) = 1/n$ for every $t \ge 0$ and $i \in [0,n-1]$.
Each $\Gamma_t$ represents the interaction that occurs
at step $t$.
Given an initial configuration $C_0$,
the \emph{execution} of protocol $P$ under $\sch$
is defined as 
$\Xi_{P}(C_0,\sch) = C_0,C_1,\dots$ such that
$C_t \stackrel{\Gamma_t}{\to} C_{t+1}$ for all $t \ge 0$. 
We denote $\Xi_{P}(C_0,\sch)$
simply by $\Xi_{P}(C_0)$ when no confusion occurs.

We address the self-stabilizing leader election problem
in this paper. For simplicity,
we give the definition of a self-stabilizing leader election protocol instead of giving the definitions of self-stabilization and the leader election problem separately.
\begin{definition}[Self-stabilizing Leader Election]
\label{def:ss-le} 
For any protocol $P$,
we say that a configuration $C$ of $P$ is \emph{safe}
if (i) exactly one agent outputs $L$ (leader)
and all other agents output $F$ (follower) in $C$,
and
(ii) at every configuration reachable from $C$,
all agents keep the same outputs as those in $C$.
A protocol $P$ is a \emph{self-stabilizing leader election (SS-LE) protocol} if $\Xi_{P}(C_0,\sch)$ reaches a safe configuration
with probability $1$ for any configuration $C_0 \in \call(P)$.
\end{definition}

We denote interaction $(u_i,u_{i+1})$ by $e_i$.
For any two sequences of interactions
$s=e_{k_0},e_{k_1},\dots,e_{k_h}$
and $s'=e_{k'_0},e_{k'_1},\dots,e_{k'_{j}}$,
we define
\[s\cdot s' = e_{k_0},e_{k_1},\dots,e_{k_h},e_{k'_0},e_{k'_1},\dots,e_{k'_{j}}.\]
That is, we use ``$\cdot$'' for the concatenation operator.
For any sequence $s$ of interactions and integer $i \ge 1$,
$s^i$ denotes the $i$-times repeating sequence of $s$:
$s^1 = s$ and $s^i = s\cdot s^{i-1}$.
For any integer $i \in [0,n-1]$
and any integer $j \ge 0$,
we define $\seqr(i,j)=e_i,e_{i+1},\dots,e_{i+j-1}$
and $\seql(i,j)=e_{i-1},e_{i-2},\dots,e_{i-j}$.
We say that a sequence $s$ of interactions has length $\ell$
if it consists of $\ell$ interactions.

\begin{definition}
Let $\gamma=e_{k_1},e_{k_2},\dots,e_{k_h}$
be a sequence of interactions.
We say that $\gamma$ occurs within $\ell$ steps when
$e_{k_1},e_{k_2},\dots,e_{k_h}$ occurs 
in this order (not necessarily in a row) 
within $\ell$ steps.
Formally,
the event
``$\gamma$ occurs within $\ell$ steps from a time step $t$''
is defined as the following event:
$\Gamma_{t_i}=e_{k_i}$ holds for all $i = 1,2,\dots,h$
for some sequence of integers 
$t \le t_1 < t_2 < \dots < t_{h} \le t+\ell-1$.
We say that from step $t$,
$\gamma$ completes at step $t+\ell-1$
if $\gamma$ occurs within $\ell$ steps
but does not occur within $\ell-1$ steps.
When the starting step $t$ is clear from the context, 
we just write ``$\gamma$ occurs within $\ell$ steps''
and ``$\gamma$ completes at step $t+\ell-1$''.
\end{definition}

\begin{lemma}
\label{lem:base} 
Let $s$ be a sequence of interactions with length $\ell$.
Then, $s$ occurs within $n\ell$ steps in expectation.
For any $c \ge 1$,
$s$ occurs within $2n(\ell + 4c\log n)$
steps with probability $1-n^{-c}$.
\end{lemma}
\begin{proof}
The first claim immediately follows from linearity of expectation and the fact that
for any $i \ge 1$, the $i$-th interaction of $s$ 
occurs with probability $1/n$ at each step.
The second claim follows from Chernoff bound (see Lemma~\ref{lem:chernoff} in Appendix). 
\end{proof}

Throughout this paper, we use ``with high probability'',
abbreviated as w.h.p., to mean with probability $1-O(1/n)$.
The set of integers $\{i,i+1,\dots,j\}$ for $i < j$
is denoted by $[i,j]$.

\begin{algorithm}[t]
\caption{$\ppl$}
\label{al:prl}
\VarAgent{
\\
\varspace $\leader \in \{0,1\}$,\\
\varspace $\id \in \{0,1\},\distb \in [0,2\psi-1],\last \in \{0,1\}$,
\\ \varspace
$\tokenB, \tokenW \in \{\bot\} \cup (([-\psi+1,-1]\cup[1,\psi]) \times \{0,1\} \times \{0,1\})$\\
\varspace $\mode \in \{\Detect,\Construct\},\ho \in [0,\cmax],\signalr \in [0,\rmax]$,\\
\hfill \texttt{//} $\cmax = \Theta(\psi) = \Theta(\log n)$\\
\varspace $\hits \in [0,\hmax], \bull \in \{0,1,2\},\shield \in \{0,1\},\signalb \in \{0,1\}$
}
\aspace

$\create()$ \tcp*{Create a leader when no leader exists.}
$\elim()$ \tcp*{Decrease \#leaders when \#leaders $\ge 2$.}
\end{algorithm}

\begin{lemma}
\label{lem:whp_expected}
For any SS-LE protocol $P$ and a function $f:\mathbb{N} \to \mathbb{R}$,
the convergence time (\ie the number of steps to reach a safe configuration) of $P$ is $O(f(n))$ steps in expectation
if the convergence time of $P$ is bounded by $f(n)$ steps w.h.p.
\end{lemma}
\begin{proof}
Let $X=\max_{C \in \call(P)} E_n(C)$,
where $E_n(C)$ is the expected convergence time when the initial configuration is $C$.
Then, we have $X \le f(n) + O(1/n) \cdot X$ 
because a self-stabilizing protocol must deal with an arbitrary initial configuration.
Solving the inequality gives $\max_{C \in \call(P)} E_n(C) = O(f(n))$.
\end{proof}
Note that this lemma depends on the assumption that $P$ is self-stabilizing.
We cannot use this kind of translation from ``w.h.p.\ time'' to ``expected time''
for non-self-stabilizing protocols.

\section{Self-stabilizing leader election on a ring}
In this section, we prove the following theorem.
\begin{theorem}
\label{theorem:main}
If an integer $\psi$ such that $\log n \le \psi\le \log n + O(1)$ is given,
there exists a self-stabilizing leader election protocol $P$ that works for any directed ring such that (i) for any $C_0 \in \call(P)$,
$\Xi_{P}(C_0,\sch)$ reaches a safe configuration within $O(n^2 \log n)$ steps
both w.h.p.\ and in expectation, and (ii) each agent uses $\polylog(n)$ states.
\end{theorem}
\noindent 
To prove the above theorem, we give an SS-LE protocol $\ppl$
that satisfies the requirements of the theorem.
Thanks to Lemma~\ref{lem:whp_expected}, we consider the convergence time of $\ppl$ only in terms of ``w.h.p.''
in the rest of this paper.
In addition, we assume $\psi \ge 2$ without loss of generality:
we can solve SS-LE in a straightforward way if $\psi = 1$
because it yields $n=2$.

The pseudocode of $\ppl$ is given in Algorithms~\ref{al:prl},~\ref{al:create},~\ref{al:token},~\ref{al:determ}, and~\ref{al:le},
which describes how two agents $l$ and $r$ update their states, \ie their variables, when they have an interaction.
Here, $l$ and $r$ represent the initiator
and the responder in the interaction, respectively.
That is, $l$ is the left neighbor of $r$,
and $r$ is the right neighbor of $l$.
We denote the value of variable $\mathtt{var}$
at agent $v\in V$ by $v.\mathtt{var}$.
Similarly, we denote the value of variable $\mathtt{var}$
in state $q\in Q$ by $q.\mathtt{var}$.
In this protocol, each agent $v \in V$ maintains
an \emph{output} variable $v.\leader \in \{0,1\}$,
according to which it determines its output.
Agent $v$ outputs $L$ when $v.\leader = 1$
and outputs $F$ when $v.\leader = 0$.
We say that $v$ is a \emph{leader} if $v.\leader = 1$; otherwise $v$ is called a \emph{follower}.
For each $u_i \in V$,
we define
the distance to \emph{the nearest left leader} 
and the distance to \emph{the nearest right leader}
of $u_i$
as 
$\distll(i) = \min \{j \ge 0 \mid u_{i-j}.\leader = 1\}$
and
$\distrl(i) = \min \{j \ge 0 \mid u_{i+j}.\leader = 1\}$,
respectively.
When there is no leader in the ring, 
we define $\distll(i)=\distrl(i)=\infty$.
By definition, $\distll(i)=\distrl(i)=0$ when $u_i$ is a leader.
Note that $\distll(i)$ and $\distrl(i)$ are not variables maintained by agents, but just notations. 

The pseudocode of the main function of $\ppl$ is shown in
Algorithm~\ref{al:prl}, which consists of two functions
$\create()$ and $\elim()$.
To design a self-stabilizing protocol,
we must consider an arbitrary initial configuration, where
there may be no leader, or there may be two or more leaders.
The goal of $\create()$ is to create a new leader 
when no leader exists in the population,
while the goal of $\elim()$ is to decrease the number of leaders to one when there are two or more leaders.
Function $\elim()$, presented in \cite{YSM21},
decreases the number of leaders to one within $O(n^2 \log n)$ steps with high probability using only a constant number of states.
Although we use $\elim()$ without any modification,
we will present $\elim()$ in Section~\ref{sec:elim} for completeness.

\subsection{Overview}
\label{sec:overview}
The difficulty of designing $\create()$ lies in detecting
the absence of a leader in a self-stabilizing fashion.
Starting from an arbitrary configuration without a leader,
the population must detect the absence of a leader
and create a new leader. 

The protocol presented in \cite{YSM21}
achieves the detection in a naive way using $O(n)$ states,
given an upper bound $N=n + O(n)$:
each agent computes the distance from the nearest left leader and 
detects the absence of a leader when the computed distance is $N$ or larger. However, $\create()$ is allowed to use only $\polylog(n)$ states, so we cannot take this approach. 
One may think that we can still use this strategy
by maintaining \emph{approximate distance} instead of the exact distance from the nearest left leader,
in order to reduce the number of states to $\polylog(n)$.
We conjecture that this does not work. 
If we try to use approximate distance, the agents sometimes compute a very large distance, which results in misdetecting the absence of a leader and creating a new leader even if the population already has a unique leader. Such an event may occur extremely rarely, however, it eventually occurs even after an extremely long expected time. So, the resulting protocol is not self-stabilizing.

The protocol presented in \cite{CC19}
detects the absence of a leader in a different way. 
Their protocol lets the population reach a safe configuration
where exactly one leader exists and
a bit-string that can prove the existence of a leader
is embedded on the ring. 
Specifically, 
a prefix of the bit-string called Thue-Morse string \cite{Thue1912} is embedded:
each agent $u_i$ has variable $u_i.B \in \{0,1\}$,
and $s_k, s_{k+1}, \dots, s_{k+n-1}$ is a prefix
of Thue-Morse string, where $s_i = u_i.B$ and $u_k$ is the unique leader. 
Although we do not write the definition of Thue-Morse string here, it has a good property called \emph{cube-freeness},
that is, it does not contain $www$ as a substring for any string $w$.
Using this property, 
the population declares the detection of the absence of a leader if and only if 
it finds a string $www$ for some $w$.
After the population reaches a safe configuration,
the cube-freeness of Thue-Morse string guarantees that
the protocol does not find such a string.
On the other hand, if there is no leader,
the agents can eventually observe $www$ for $w=s_{i},s_{i+1},\dots,s_{i+n-1}$ for some $i$, letting them detect the absence of a leader. Chen and Chen \cite{CC19} implement this idea with $O(1)$ states per agent.
However, implementing this idea is not easy; consequently,
their protocol requires a super-exponential number of steps in expectation and w.h.p.

\begin{figure}[t]
\centering
\includegraphics[width= 0.85 \linewidth,clip]{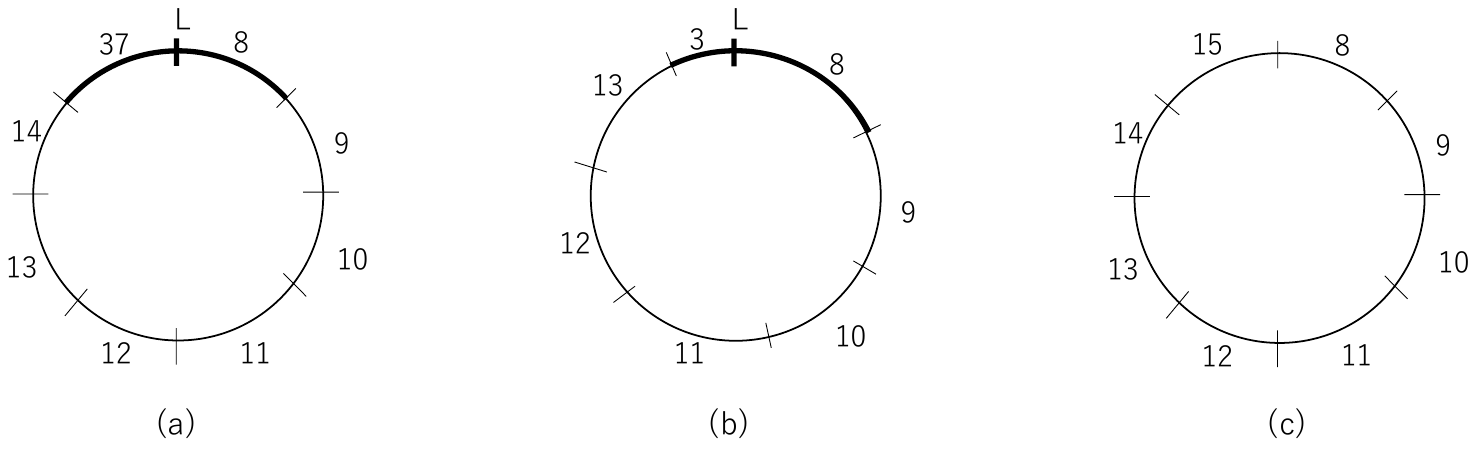}
\caption{
Example of embedding of segment IDs on the ring.
$L$ represents a leader.
}
\label{zu:segments}
\end{figure}

We embed a much simpler string on the ring to 
achieve much faster convergence time, \ie $O(n^2 \log n)$ steps,
at the cost of increasing the number of states
from $O(1)$ to $\polylog (n)$.
Each $u_i$ maintains two variables $u_i.\distb \in [0,2\psi-1]$ and $u_i.\id \in \{0,1\}$ for the string embedding.
When the embedding is complete,
\begin{align}
\label{eq:dist}
u_i.\distb = 
\begin{cases}
0 & \text{if } u_i \text{ is a leader} \\
(u_{i-1}.\distb + 1) \bmod{2\psi}& \text{otherwise}
\end{cases} 
\end{align}
must hold.
We call an agent $u_i$ that satisfies
$u_i.\distb \in \{0,\psi\}$ a \emph{border}.
We say that a sequence of agents $u_i,u_{i+1},\dots,u_{i+\ell-1}$ is a \emph{segment} when $u_i$ and $u_{i+\ell}$ are borders and $u_{i+1}, u_{i+2},\dots,u_{i+\ell-1}$ are not borders.
Let $b_i = u_i.\id$.
We define the ID of a segment $S = u_i,u_{i+1},\dots,u_{i+\ell-1}$
as $\iota(S) = \sum_{j=0}^{\ell-1} b_{i+j} \cdot 2^{j}$,
that is, $\iota(S)$ is the integer that corresponds to
a bit string $b_{i+\ell-1} b_{i+\ell-2} \dots b_{i}$
in the base-2 numeral system.
For a segment $S = u_i,u_{i+1},\dots,u_{i+\ell-1}$,
we denote the previous segment $u_j,u_{j+1},\dots,u_{i-1}$ by $\prev(S)$.
When the embedding is complete, 
a segment $S = u_i,u_{i+1},\dots,u_{i+\ell-1}$ must satisfy

\begin{align}
\label{eq:seg}
\iota(S) = (\iota(\prev(S))+1) \bmod{2^\psi}
\hspace{0.5cm} \text{or} \hspace{0.5cm}
u_i \text{ or } u_{i+\ell} \text{ is a leader.}
\end{align}
We say that
a configuration $C$ is \emph{perfect}
if no violation occurs against \eqref{eq:dist} or \eqref{eq:seg}
in $C$.
For any $n$ and $\psi$, there exists a perfect configuration with one leader.
See examples in Figure~\ref{zu:segments} (a) and (b).
The segment IDs increase one by one in the clockwise direction starting with the segment that contains a leader, while we do not need to care about the IDs of the first and the last segments,
which are drawn in bold curves in Figure~\ref{zu:segments}.
Importantly, a perfect assignment requires the existence of a leader.
That is, every configuration containing no leader violates
\eqref{eq:dist} or \eqref{eq:seg}.
For example,
in Figure~\ref{zu:segments} (c) with $\psi = 7$, the segment with ID 8 violates condition \eqref{eq:seg} because $8 \neq (15+1) \bmod{2^7}$. 
This claim can be easily proved
from the fact that $2^\psi \ge 2^{\log n} = n$ and $\psi \ge 2$, as follows.

\begin{lemma}
\label{lem:impossible} 
A configuration without a leader is not perfect.
\end{lemma}

\begin{proof}
Assume for contradiction
that there is a perfect configuration $C$ that does not have a leader.
Since \eqref{eq:dist} holds for all $u_i$ and there is no leader in $C$,
the length of all segments is $\psi$, thus the number of segments is exactly $n/\psi < n \leq 2^\psi$ in $C$.
This yields that there is at least one segment $S$ that violates \eqref{eq:seg},
which contradicts the assumption.
\end{proof}

\setcounter{AlgoLine}{2}
\begin{algorithm}[t]
\caption{$\create()$\ \ \ $l$ is the initiator and $r$ is the responder of an interaction.}
\label{al:create}
$\determ()$\;

$\tmp \gets \begin{cases}
	0 & r.\leader = 1\\		 
	l.\distb + 1 \bmod{2\psi} & \text{otherwise}
	\end{cases}$\;
\tcp*{$\tmp$ is a temporary variable}

\If{$r.\mode = \Detect \land \tmp \neq r.\distb$}{
$(r.\leader,r.\bull,r.\shield,r.\signalb) \gets (1,2,1,0)$\;
\tcp*{create a leader.}
}
\If{$r.\mode = \Construct$}{
$r.\distb \gets \tmp$
}
$l.\last \gets
\begin{cases}
1 &r.\leader = 1\\ 
0 & r.\leader = 0 \land r.\distb \in \{0,\psi\} \\
r.\last & \text{otherwise} 
\end{cases}$\;
$\MT(\tokenB,0)$\;
$\MT(\tokenW,\psi)$\;
\end{algorithm}

The basic structure of $\create()$, whose pseudocode is shown in Algorithm~\ref{al:create}, is as follows.
\begin{enumerate}
 \item Each agent $u_i$ maintains a variable $u_i.\mode \in \{\Detect,\Construct\}$. We say that $u_i$ is in the detection mode if $u_i.\mode = \Detect$; otherwise $u_i$ is in the construction mode.
Function $\determ()$, called at Line~3, determines which mode the agents in the population should be in. As we will see later,
$\determ()$ guarantees that
\begin{itemize}
 \item once a leader exists in the population, all agents will enter the construction mode within $O(n^2 \log n)$ steps and keep the construction mode
 in the next $c n^2 \log n$ steps for arbitrarily large constant $c$ 
 w.h.p.\footnote{
 Actually, this constant $c$ depends on the parameter $\cmax$. We can increase $c$ arbitrarily by increasing $\cmax$
 so that the whole population stays in the construction mode for a sufficiently long time.
 },
and 
 \item if there is no leader, a new leader is created or all the agents will enter the detection mode within $O(n^2 \log n)$ steps  w.h.p., after which no agent goes back to the construction mode until a new leader is created.
\end{itemize}
 \item The agents in the construction mode try to let the population reach a perfect configuration by updating variables $\distb$ and $\id$. If all agents are in the construction mode and there is one leader, 
they will, within $O(n^2 \log n)$ steps w.h.p., make a perfect configuration from which
no agent changes $\distb$ and $\id$. 
 \item The agents in the detection mode try to detect the imperfection of the current configuration. If all agents are in the detection mode and there is no leader, they will detect the imperfection and create a new leader within $O(n^2 \log n)$ steps w.h.p. 
\end{enumerate}

Thus, the combination of $\create()$ and $\elim()$
lets the population reach, within $O(n^2 \log n)$ steps w.h.p.,
a safe configuration such that
there is exactly one leader
and no leader will be created in the following execution.
Function $\elim()$ may kill (i.e., remove) a leader even if 
it is the unique leader in the population.
This is due to the inconsistency of variables that might occur in some configurations. (Recall that we must consider an arbitrary initial configuration to design a self-stabilizing protocol.)
However, as we will see in Section~\ref{sec:correctness}, $\create()$ and $\elim()$ will remove this inconsistency
within $O(n^2)$ steps w.h.p. 
Thereafter, $\elim()$ will never kill the last leader in the population, i.e., it will remove
a leader only when there are two or more leaders.
If there is no leader,
$\determ()$ lets all agents enter the detection mode, after which 
$\create()$ lets the agents detect the imperfection of the current configuration and create a new leader (not necessarily a single leader),
which requires $O(n^2 \log n)$ steps w.h.p.
Once a leader appears in the population,
$\determ()$ lets all agents enter the construction mode
within $O(n^2)$ steps and keep the construction mode for a sufficiently large $\Theta(n^2 \log n)$ steps w.h.p.
As mentioned above, $\elim()$ decreases the number of leaders to one within $O(n^2 \log n)$ steps w.h.p.
Thus, during the period where all agents are in the construction mode,
$\elim()$ elects a unique leader, and $\create()$ updates $\distb$ and $\id$ so that the population reaches a configuration $C$ such that (i) $C$ is perfect, and (ii) no agent changes $\distb$ or $\id$ from $C$
unless a new leader is created.
Thereafter, no agent creates a new leader even if some agent goes back to the detection mode.
Thus, the population keeps the unique leader forever.

The rest of this section is organized as follows.
In Section~\ref{sec:create}, we explain how $\create()$ constructs a perfect configuration when the agents are in the construction mode
and how it detects the imperfection of the current configuration when the agents are in the detection mode and
there is no leader.
In Section~\ref{sec:determ}, we explain how $\determ()$
determines the mode of the agents and how it achieves the desirable behavior mentioned above. 
In Section~\ref{sec:elim}, we explain how $\elim()$ given by \cite{YSM21}
reduces the number of leaders to one for completeness.

\setcounter{AlgoLine}{11}
\begin{algorithm}
\caption{$\MT(\token,d)$\ \ \ $l$ is the initiator and $r$ is the responder of an interaction.}
\label{al:token}
\Macros{\\
\varspace
$\target(v,d) = (v.\distb + v.\token[1] + d) \bmod 2\psi$
\\
\varspace $\valid(v,d) = 
\begin{cases}
1 & v.\token[1] > 0
\land \target(v,d) \in [\psi,2\psi-1]
\\
1 & v.\token[1] < 0
\land \target(v,d) \in [1,\psi-1]
\\
0 & \text{otherwise}
\end{cases}$\\
}
\aspace

$v.\token \gets \bot$ for each $v \in \{l,r\} \text{ s.t.~} v.\token \neq \bot \land \valid(v,d) = 0$\;
\tcp*{remove all invalid tokens}
\If{$l.\last = r.\last = 0$}{
\If{$l.\distb = d \land l.\token = \bot$}{
$l.\token \gets (\psi, 1 - l.\id, l.\id)$ \tcp*{create a new token}
}
\If{$l.\token \neq \bot \land r.\token \neq \bot $}{
$l.\token \gets \bot$ \tcp*{two tokens are merged}
}

\uIf(\tcp*[f]{token hits target}){$l.\token \neq \bot \land l.\token[1] = 1$}{
\uIf{$r.\mode = \Detect \land l.\token[2] \neq r.\id$}{
$(r.\leader,r.\bull,r.\shield,r.\signalb) \gets (1,2,1,0)$\;
\tcp*{create a leader}
}
\ElseIf{$r.\mode = \Construct$}{
 $r.\id \gets l.\token[2]$
}
 $r.\token \gets (1-\psi,l.\token[2],l.\token[3])$\;
 $l.\token \gets \bot$\;
}
\uElseIf{$l.\token \neq \bot \land l.\token[1] \ge 2$}{
$r.\token \gets (l.\token[1]-1,l.\token[2],l.\token[3])$\;
$l.\token \gets \bot$\;
}

\uElseIf(\tcp*[f]{token hits target}){$r.\token \neq \bot \land r.\token[1] = -1$}{
$l.\token \gets
\begin{cases}
 (\psi,1-l.\id,l.\id)& r.\token[3] = 1\\
 (\psi,l.\id,0)& r.\token[3] = 0
\end{cases}$\;
$r.\token \gets \bot$

}
\ElseIf{$r.\token \neq \bot \land r.\token[1] \le -2$}{
$l.\token \gets (r.\token[1]+1,r.\token[2],r.\token[3])$\;
$r.\token \gets \bot$\;
}
}
\end{algorithm}

\subsection{Construction and detection}
\label{sec:create}

In this section, we explain how $\create()$ guarantees that
\begin{itemize}
 \item when all agents are in the construction mode and there is exactly one leader,
the population reaches a perfect configuration within $O(n^2 \log n)$ steps w.h.p., and
 \item when all agents are in the detection mode and there is no leader,
the imperfection of the current configuration is detected and
a new leader is created within $O(n^2 \log n)$ steps w.h.p.
\end{itemize}
Thus, we do not consider the case where there are two or more leaders in the population throughout this section.

The pseudocode of $\create()$, shown in Algorithm~\ref{al:create},
consists of three parts:
the first part manages variable $\mode$ by invoking $\determ()$
(Line~3),
the second part manages variables $\distb$ and $\last$ (Lines~4--9),
and the third part manages variable $\id$ by invoking a function $\MT()$ (Lines~10--11).
The first part is left to Section~\ref{sec:determ}.
The second and third parts guarantee the above specification
by updating and checking variables $\distb$, $\last$, and $\id$.

A configuration is said to be \emph{good} if
\begin{itemize}
    \item no agent violates the first condition of perfection (i.e., equation \eqref{eq:dist}), and
    \item $v.\last = 1$ holds if and only if $v$ belongs to the last segment, i.e.,
    the segment $u_i, u_{i+1}, \dots, u_{i+\ell -1}$ such that $u_{i+\ell}$ is a leader.
\end{itemize}
The goal of the second part is as follows:
(i) if all agents are in the construction mode and there is exactly one leader, 
the population enters a good configuration,
and (ii) if all agents are in the detection mode and there is no leader, 
the population enters a good configuration or creates a new leader.

This goal can be achieved as follows.
When $u_j$ and $u_{j+1}$ interact,
the right agent $u_{j+1}$ tries to compute its distance to the nearest
left leader modulo $2\psi$
in a temporary variable $\tmp$ as follows:
$\tmp \gets 0$ if $r$ is a leader, and
$\tmp \gets l.\distb + 1 \bmod 2\psi$ otherwise.
If $u_{j+1}$ is in the construction mode,
it simply assigns the computed value to $u_{j+1}.\distb$.
If $u_{j+1}$ is in the detection mode and
$\tmp \neq u_{j+1}.\distb$,
this implies that the current configuration is not perfect.
In this case, $u_{j+1}$ creates a new leader.
(For now, ignore the variables $\bull$, $\shield$, and $\signalb$ at Line~6.
We explain these variables in Section~\ref{sec:elim}.)
The variable $v.\last \in \{0,1\}$ indicates whether
agent $v$ belongs to the last segment, i.e.,
the segment $u_i, u_{i+1}, \dots, u_{i+\ell -1}$
such that $u_{i+\ell}$ is a leader.
At each interaction, the left agent updates $\last$ at Line~9,
regardless of its mode.

Consider that interaction sequence $e_{i-1} \cdot \seqr(i,n) \cdot \seql(i,n)$ occurs, during which
$u_i$ is always a unique leader and all agents are in the construction mode.
Then, the population reaches a configuration where 
no agent violates the first condition of perfection,
\ie \eqref{eq:dist},
and the agents in the last segment have $\last = 1$
and the other agents have $\last = 0$.

Consider that $\seqr(i,n)\cdot\seql(i,n)$ occurs for some $i \in [0,n-1]$,
during which 
there is no leader and all agents are in the detection mode.
If the population has some inconsistency in variable $\distb$,
\ie $u_{j+1}.\distb \neq u_j.\distb + 1 \bmod 2\psi$ and $u_j.\leader=0$ for some $j$,
at least one agent detects the inconsistency and creates a leader in this period.
Otherwise, the population reaches a configuration where 
$v.\last = 0$ holds for all agents $v$ in this period.

In both cases, $\seqr(i,n) \cdot \seql(i,n)$ occurs within $O(n^2)$ steps w.h.p.\ by Lemma~\ref{lem:base}.
So, the above goal is achieved within $O(n^2)$ steps w.h.p.

Thus, in the rest of Section~\ref{sec:create},
we explain how the third part works
starting from any good configuration.
Again, we consider only two cases in this section (Section~\ref{sec:create}): (i) there is exactly one leader,
and (ii) there is no leader. 
Without loss of generality,
we assume that $u_0$ is the unique leader in the first case,
while we assume $u_0.\distb = 0$ in the second case.
Then, we have $\zeta \overset{\text{def}}{=} \lceil n/\psi \rceil$ segments
$S_0, S_1, \dots, S_{\zeta-1}$ in the population
where $S_i = u_{i\psi},u_{i\psi + 1},\dots,u_{i\psi + \psi -1}$ for $i \in [0,\zeta-2]$
and $S_{\zeta-1} = u_{(\zeta-1)\psi},u_{(\zeta-1)\psi+1},\dots,u_{n-1}$. We say that a segment $S_i$ is \emph{black}
if $i$ is even; otherwise $S_i$ is \emph{white}.
We call the border agent in a black (resp.~white) segment
a \emph{black border} (resp.~\emph{white border}).

\begin{figure}[t]
\centering
\includegraphics[width= 0.7 \linewidth,clip]{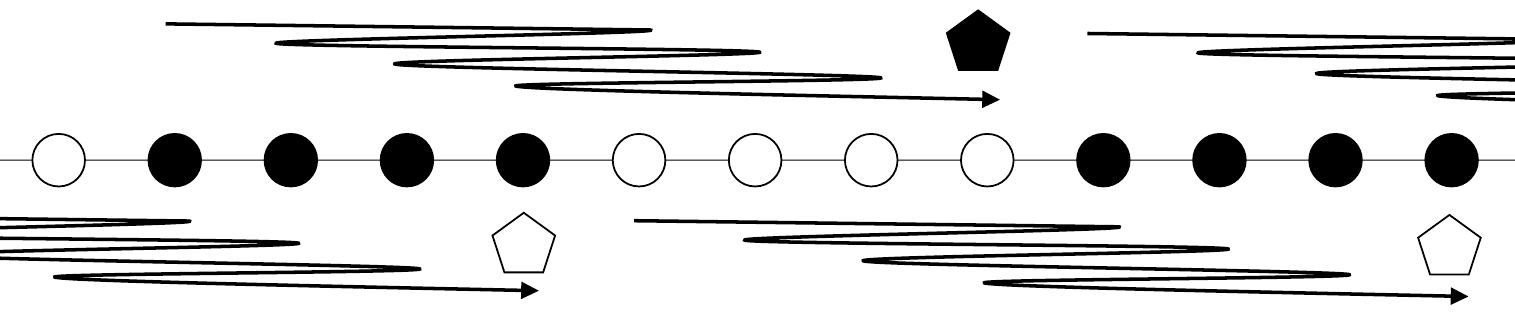}
\caption{
The trajectory of black and white tokens. ($\psi = 4$.)
}
\label{zu:tokens}
\end{figure}

The goal of the third part is as follows:
for each $i \in [0,\zeta-3]$,
it embeds $(\iota(S_i)+1) \bmod 2^\psi$ on variable $\id$ on the agents in $S_{i+1}$ to satisfy $\iota(S_{i+1}) = (\iota(S_i)+1) \bmod 2^\psi$
in the construction mode, 
while it creates a new leader when it finds $\iota(S_{i+1}) \neq (\iota(S_i)+1) \bmod 2^\psi$ in the detection mode.
To achieve this goal,
we use two kinds of tokens, black tokens and white tokens.
For each $i \in [0,\zeta-2]$,
the border agent $u_{i\psi}$ in $S_i$ creates a black (resp.~white) token if it is black (resp.~white), and the generated token moves back and forth in the trajectory shown in Figure~\ref{zu:tokens}, conveying two bits $b'$ and $b''$.
These two bits are used to compute $(\iota(S_i)+1) \bmod 2^\psi$:
$b'$ maintains the binary value that will be set to the target index
and $b''$ maintains the carry flag.
Specifically,
the agents in two segments $S_i$ and $S_{i+1}$
move the token in the following steps:
\begin{enumerate}
 \item When $u_{i\psi}$ creates a token, 
it initializes $b'$ and $b''$ as
\[
(b',b'') \gets (1-u_{i\psi}.\id,u_{i\psi}.\id),
\]
and sets the target index $T = \psi$.
 \item The token moves right toward $u_r = u_{i\psi + T}$.
 \item  The token arrives at $u_r$ eventually.
If $u_{r}$ is in the construction mode at that time,
it copies $b'$ to $u_{r}.\id$.
Otherwise,
it does not update $u_{r}.\id$.
Instead, it checks whether $b' = u_{r}.\id$ holds.
If it does not hold, $u_{r}$ becomes a leader.
 \item The token disappears if $T = 2\psi - 1$, \ie it has already reached the final destination $u_{i\psi + 2\psi -1}$. Otherwise, it begins to move left toward $u_l = u_{i\psi+T-\psi + 1}$.
 \item  The token moves left toward $u_l$.
 \item  The token arrives at $u_l$ eventually.
 It updates $b'$ and $b''$ as 
\[
(b',b'') \gets \begin{cases}
(1-u_l.\id,u_l.\id) & b'' = 1 \\
(u_l.\id,0) & b'' = 0,
 \end{cases}
\]
and increments the target index $T$ by one
 and goes back to Step 2.
\end{enumerate}

This behavior of the token obviously achieves the above goal. 
We realize this behavior with two variables $\tokenB, \tokenW \in \{\bot\} \cup (([-\psi+1,-1]\cup[1,\psi]) \times \{0,1\} \times \{0,1\})$, which represent black and white tokens, respectively.
For any agent $u_i$, $u_i.\tokenB = \bot$ (resp.~$u_i.\tokenW = \bot$) means that $u_i$ is not carrying a black (resp.~white) token currently.
In the following, we only explain how the agents realize the behavior of a black token with variable $\tokenB$. The behavior of a white token is realized in completely the same way.

If $u_i.\tokenB \neq \bot$, the black token consists of three elements. The first element $u_i.\tokenB[1] \in [-\psi+1,-1]\cup[1,\psi]$ represents the relative position of the target.
When $u_i.\tokenB[1] \ge 1$, the black token at $u_i$ is moving right toward $u_{i+u_i.\tokenB[1]}$.
When $u_i.\tokenB[1] \le -1$, the token is moving left toward 
$u_{i+u_i.\tokenB[1]}$.
The second element $u_i.\tokenB[2] \in \{0,1\}$ and the third element $u_i.\tokenB[3] \in \{0,1\}$ respectively correspond to the two bits $b'$ and $b''$ that were
used to describe the behavior specified by Steps 1--6. 
Each step $1,2,\dots,6$ is realized by the following lines
at Algorithm~\ref{al:token}:
\begin{itemize}
 \item Step 1: Lines~14--15,
 \item Step 2: Lines~26--27,
 \item Step 3: Lines~19--22,
 \item Step 4: Lines~12 and 23--24,
 \item Step 5: Lines~32--33,
 \item Step 6: Lines~29--30.
\end{itemize}
Note that in the pseudocode, we use $d=0$ and $\token=\tokenB$ for black tokens,
whereas we use $d=\psi$ and $\token=\tokenW$ for white tokens
(see Lines~10 and~11 in Algorithm~\ref{al:create}).
The above implementation of the behavior of the tokens is straightforward and requires no explanation except for Line 12 and Lines~16--17.

Line~12 is devoted to deleting \emph{invalid tokens}.
Roughly speaking, a token is invalid if it is outside its trajectory.
Formally, we define valid and invalid tokens as follows.

\begin{definition}
\label{def:invalid}
\begin{em}
A token located at agent $u_j$ is \emph{valid}
if
\begin{align*}
&u_j.\token[1] > 0  
\land
(u_j.\distb + u_j.\token[1] + d) \bmod 2\psi \in [\psi,2\psi-1],\\
&\text{or } u_j.\token[1] < 0
\land 
(u_j.\distb + u_j.\token[1] + d) \bmod 2\psi \in [1,\psi-1],
\end{align*}
where $d=0$ if the token is black, and $d=\psi$
otherwise.
A token is \emph{invalid} if it is not valid.
\end{em}
\end{definition}

\noindent
A valid token never becomes invalid before it reaches its final destination $u_\ell$,
where $\ell = (d+2\psi-1) \bmod 2\psi$.
When the token reaches $u_\ell$, it becomes invalid because
$u_\ell.\token[1]$ is set to $1-\psi<0$, by which
\[
(u_\ell.\distb + u_\ell.\token[1] + d) \bmod 2\psi = \psi
\]
holds.
Therefore, the token disappears the next time $u_\ell$ has an interaction by Line~12.
Invalid tokens may exist anywhere in an initial configuration,
which the adversary may choose arbitrarily.
However, this does not matter for the following reason.
\begin{note}
\label{note:invalid}
We can ignore invalid tokens when analyzing the correctness and time complexities of $\ppl$
because Line~12 deletes any invalid token whenever the hosting agent has an interaction,
and Line~12 is executed before any other command involving tokens.
\end{note}

Lines~16--17 deletes the left token when two black tokens meet. This destruction does not cause live-lock because the rightmost valid black token in the segments $S_i$ and $S_{i+1}$ is never killed until it reaches the final destination $u_{i\psi+2\psi-1}$.

\begin{note}
All commands involving tokens, except for Line~12, which is devoted to deleting invalid tokens, are executed 
only when two agents with $\last = 0$ interact (see Line~13).
Therefore, as long as the population remains in good configurations, 
no agent detects
\[
\iota(S_{\zeta-1}) \neq \iota(S_{\zeta-2}) + 1 \bmod 2^\psi,
\text{ or} \quad 
\iota(S_{0}) \neq \iota(S_{\zeta-1}) + 1 \bmod 2^\psi.
\]
Hence, such inconsistencies, which we intentionally ignore, never trigger the creation of a new leader.
\end{note}

The length of the trajectory of a token is exactly
$(\psi + \psi -1)\cdot (\psi - 1) + \psi = 2\psi^2-2\psi + 1$.
Thus, under the assumption that variable $\distb$ is correctly computed,
every token moves at most $2\psi^2-2\psi + 1$ times.
\begin{definition}
\label{def:token_complete} 
We say that a token \emph{completes its trajectory}
when it moves $2\psi^2-2\psi + 1$ times.
\end{definition}

\begin{lemma}
\label{lem:complete} 
Let $k \in [0,n-1]$ and $d \in \{0,\psi\}$.
Let $\calC_d$ be the set of all configurations where 
$u_{k+i}.\distb = i + d \bmod 2\psi$ and $u_{k+i}.\last = 0$ hold
for all $i \in [0,2\psi-1]$.
Consider execution $\Xi_{\ppl}(C_0,\sch) = C_0,C_1,\dots$.
If $C_0,C_1,\dots,C_{\ell} \in \calC_d$ holds
and interaction sequence $(\seqr(k,2\psi-1)\cdot \seql(k+2\psi-1,2\psi-1))^{2\psi}$ occurs within the first $\ell$ steps,
there exists at least one token $z$ such that
(i) $z$ does not exist in $C_0$,
(ii) $z$ is generated by $u_k$,
and (iii) $z$ completes its trajectory 
in $C_0,C_1,\dots,C_{\ell}$.
\end{lemma}

\begin{proof}
We prove the lemma only for the case $d=0$.
The case $d=\psi$ can be handled similarly. 
Since $d=0$, $u_k$ is a black border.
Let $t\ (\le \ell)$ be the time step at which the first 
$(\seqr(k,2\psi-1)\cdot \seql(k+2\psi-1,2\psi-1))^\psi$ completes.
Then, all black tokens that exist at 
$u_k,u_{k+1},\dots,u_{k+2\psi-1}$ in $C_0$ disappear
during $C_0,C_1,\dots,C_t$.
If there exists one or more black tokens
at $u_k,u_{k+1},\dots,u_{k+2\psi-1}$ in $C_t$,
let $z$ be the rightmost valid black token among them. 
Otherwise, let $z$ be the first token generated by $u_k$ after step $t$.
By definition, $z$ does not exist in $C_0$ and is generated by $u_k$.
Moreover, by the definition of valid tokens, every valid black token outside the region
$u_k,u_{k+1},\dots,u_{k+2\psi-1}$
never enters this region.
Hence, by Note~\ref{note:invalid}, 
$z$ completes its trajectory in $C_0,C_{1},\dots,C_{\ell}$.
\end{proof}

Consider the case that there exists exactly one leader
and all agents are in the construction mode.
Then, the third part (\ie the movement of black and white tokens)
correctly constructs the segment IDs so that $\iota(S_{i+1}) = (\iota(S_i)+1) \bmod 2^\psi$ holds for all $i \in [0,\zeta-3]$,
\ie lets the population reach a perfect configuration,
while $(\seqr(0,2\psi-1)\cdot \seql(2\psi-1,2\psi-1))^{2\psi} \cdot (\seqr(\psi,2\psi-1)\cdot \seql(3\psi-1,2\psi-1))^{2\psi} \cdot \dots \cdot (\seqr((\zeta-3)\psi,2\psi-1)\cdot \seql((\zeta-1) \psi - 1,2\psi-1))^{2\psi}$ occurs, by Lemma~\ref{lem:complete}.
This interaction sequence has length at most $(4\psi-2) \cdot (2\psi) \cdot (\zeta - 1) \le 8n \psi$, thus occurs within at most $O(n^2 \psi) = O(n^2 \log n)$ steps w.h.p.\ by Lemma~\ref{lem:base}.

Consider the case that there is no leader 
and all agents are in the detection mode.
Then, under the aforementioned assumption, 
all segments $S_0, S_1, \dots, S_{\zeta-1}$ have length $\psi$
and all agents have $\last = 0$.
Thus, by Lemma~\ref{lem:impossible}, 
there exists a segment $S_i$ such that $\iota(S_{(i+1)}) \neq (\iota(S_i)+1) \bmod 2^\psi$.
Then, the third part detects the inconsistency and creates a new leader
while $(\seqr(i\psi,2\psi-1)\cdot \seql((i+2)\psi-1,2\psi-1))^{2\psi}$ occurs. This interaction sequence occurs
within at most $O(n\psi^2) = O(n \log ^2 n)$ steps w.h.p.\ by Lemma~\ref{lem:base}.

\subsection{Mode determination}
\label{sec:determ}
To describe the goal of $\determ()$ properly,
we define three sets of configurations,
$\lexist$, $\lzero$ and $\cnz$.
We define $\lexist \subseteq \call(\ppl)$ 
as the set of configurations
with at least one leader.
We define $\lzero = \call(\ppl) \setminus \lexist$.
We define $\cnz \subseteq \lexist$ as the set of all configurations
from which no configuration in $\lzero$ is reachable,
\ie in any execution of $\ppl$ starting from a configuration in $\cnz$,
the last leader is never killed.
We will prove in Section~\ref{sec:correctness} that 
$\cnz \neq \emptyset$ holds, and
an execution of $\ppl$ reaches a configuration in $\cnz$ within $O(n^2 \log n)$ steps
 w.h.p.\ starting from any configuration.
The goal of this section is to give $\determ()$ such that the following two lemmas hold
for a parameter $\cmax = c_1 \psi = \Theta(\log n)$ with a sufficiently large constant $c_1$.
We assume $c_1 \ge 48$, \ie $\cmax \ge 48\psi$.

\begin{lemma}
\label{lem:keep_construction}
There exists a set $\cmid$ of configurations such that
(i) an execution of $\ppl$ starting from any configuration in $\cnz$
reaches a configuration in $\cmid$ within $O(n^2)$ steps w.h.p., 
and (ii) all agents are in the construction mode
in the first $\Omega(\cmax \cdot n^2)$ steps w.h.p.\ in an execution of $\ppl$ starting from any configuration in $\cmid$.
\end{lemma}

\begin{lemma}
\label{lem:enter_detection} 
Let $\cdet$ be the set of all configurations $C \in \lzero$ such that,
in any execution of $\ppl$ starting from $C$,
all agents remain in the detection mode until a leader is created.
Let $C_0 \in \call(\ppl)$.
Execution $\Xi_{\ppl}(C_0,\sch)$ reaches a configuration
in $\cdet \cup \lexist$
within $O(n^2 \log n)$ steps w.h.p.
\end{lemma}

\setcounter{AlgoLine}{33}
\begin{algorithm}[t]
\caption{$\determ()$\ \ \ $l$ is the initiator and $r$ is the responder of an interaction.}
\label{al:determ}
\If{$l.\leader = 1$}{
$l.\signalr \gets \rmax$\;
}
$l.\hits \gets 0$\;
$r.\hits \gets \min(r.\hits + 1,\hmax)$\;
\uIf{$l.\signalr > 0 \lor r.\signalr > 0$}{
$(l.\ho, r.\ho) \gets (0,0)$\;
\If{$l.\signalr \ge r.\signalr$}{
$(r.\signalr, r.\hits) \gets (l.\signalr,0)$\\
\tcp*[f]{the left signal absorbs the right signal}
}
$l.\signalr \gets 0$\;
\If{$r.\hits=\hmax$}{
$r.\signalr \gets r.\signalr -1$\;
$r.\hits \gets 0$\;
}
}
\ElseIf{$r.\hits=\hmax$}{
$r.\ho \gets \min(r.\ho + 1,\cmax)$\;
$r.\hits \gets 0$\;
}
\ForEach{$v \in \{l,r\}$}{
$
v.\mode \gets
\begin{cases}
 \Detect & v.\ho = \cmax\\
 \Construct & v.\ho < \cmax
\end{cases}
$
}
\end{algorithm}

We present $\determ()$ in the first half of Section~\ref{sec:determ}
and prove Lemmas~\ref{lem:keep_construction} and~\ref{lem:enter_detection}
in the second half. The pseudocode of $\determ()$ is given in
Algorithm~\ref{al:determ}.

We design $\determ()$ with a variable $\ho \in [0,\cmax]$.
We use this variable as a barometer of the absence of a leader:
high value of $u_i.\ho$ indicates a high chance that there is no leader. 
The mode (detection or construction) of an agent $u_i$ is uniquely determined
by this variable:
$u_i.\mode = \Detect$ if and only if $u_i.\ho = \cmax$ (Lines~49--50).
The basic idea is as follows.
A leader always generates a \emph{resetting signal}, which indicates that there is a leader in the population.
This signal moves in the ring in the clockwise direction, \ie from left to right,
resetting the clocks of the agents it visits.
To implement the resetting signal, each agent maintains a variable $\signalr \in [0,\cmax]$. The value of $u_i.\signalr$ represents
the TTL (Time To Live) of the signal that $u_i$ carries,
and $u_i.\signalr = 0$ means that $u_i$ does not carry a resetting signal.
A leader generates a new signal when it interacts with its right neighbor (Lines~34--35).
Each time an agent with a signal interacts with its right neighbor,
the signal moves from left to right (Line~40--42).
If the right neighbor also carries a signal, 
the two signals are merged at the right neighbor and the merged signal will get the higher TTL.
Each agent resets the clock to zero when it observes the signal
(Line~39).
Thus, the existence of a leader prevents the clock of each agent from reaching $\cmax$,
while
all agents increase their clocks to $\cmax$
and thus will enter the detection mode 
in $O(n^2 \log n)$ steps w.h.p.\ when there is no leader: as we will see later,
each $u_i$ requires $\Theta(n^2 \log n)$ steps with probability $1-O(1/n^2)$ to increase its clock by $\cmax/2$.

We have to decrease the TTL of resetting signals in some way because
we must remove all signals from the population
when no leader exists. 
On the other hand, we require that
a newly generated signal traverses the whole ring (\ie moves right $n$ times) w.h.p.\ before it disappears,
which is necessary to reset the clocks of all agents.
To achieve this with $\polylog(n)$ states,
we implement signals using the lottery game \cite{AAE+17}.
Each agent maintains a variable $\hits \in [0,\psi]$.
Agent $u_i$ increases $u_i.\hits$ by one when it interacts with its left neighbor (Line~37),
while it resets $u_i.\hits$ when it interacts with its right neighbor (Line~36).
The uniform randomness of the scheduler assures that 
$u_i$ interacts with its right neighbor with probability $1/2$
each time $u_i$ has an interaction.
If a signal at agent $u_i$ observes $u_i.\hits = \psi$, 
it decreases its TTL by one
and $u_i.\hits$ is reset to zero (Lines~43--45).
Consider that $u_i$ and $u_{i+1}$ have an interaction when both have resetting signals,
thus those signals are merged. 
Then, we say that $u_i$'s signal \emph{absorbs} $u_{i+1}$'s signal
if $u_i.\signalr \ge u_{i+1}.\signalr$;
otherwise we say that $u_{i+1}$'s signal absorbs $u_i$'s signal.
To simplify the analysis, we reset $u_{i+1}.\hits$ to zero
in the former case, \ie when $u_i$'s signal moves right absorbing $u_{i+1}$'s signal (Lines~40--42).
Note that a signal at $u_i$ moves right if and only if $u_i$ interacts with its right neighbor. 
Thus, a signal decreases its TTL
if and only if
it observes $\psi$ interactions without moving right.
Thus, every signal requires $\Theta(\cmax \cdot 2^\psi) = \Theta(n \log n)$ interactions (\ie $\Theta(n^2 \log n)$ steps) with probability $1-O(1/n^2)$ before it is absorbed or disappears (\ie decreases its TTL to zero).\footnote{
This requires careful analysis.
We will prove this fact as Lemma~\ref{lem:absorbed} later.
}
Since a signal moves right $n$ times in $O(n^2)$ steps w.h.p.\ by Lemma~\ref{lem:base}, the signal can visit all agents w.h.p.
On the other hand, when no leader exists,
the population will reach a configuration with no signals
in $O(n^2 \log n)$ steps 
 w.h.p.\ because no signal is newly generated.

We use completely the same mechanism to increase a variable $\ho$.
Each time $u_i$ observes $u_i.\hits = \psi$, it increases $u_i.\ho$ by one (Lines~46--48). Thus, when no signal exists in the population,
each $u_i$ requires $\Theta(n^2 \log n)$ steps with probability $1-O(1/n^2)$ to increase its clock by $\cmax$.
Therefore, when no leader exists,
a new leader is created or all agents enter the detection mode
within $O(n^2 \log n)$ steps w.h.p.
On the other hand, when there is at least one leader,
the population reaches a configuration where the clocks of all agents are at most $\cmax/2$
within $\Theta(n^2)$ steps w.h.p., and all agents remain in the construction mode (i.e., no agent increases its clock to $\cmax$) in the next $\Theta(n\cmax \cdot 2^\psi)=\Theta(n^2 \log n)$ steps w.h.p.

 To prove Lemmas~\ref{lem:keep_construction} and~\ref{lem:enter_detection},
we define the lottery game formally.

\begin{definition}
\label{def:lottery}
Let $k$ and $\ell$ be positive integers.
Consider that a player makes independent coin flips repeatedly.
The result of each flip is \emph{head} with probability $1/2$
and \emph{tail} with probability $1/2$.
One round of \emph{the lottery game} ends
each time the player sees a tail or consecutive $k$ heads.
If a round ends in the former case (\ie the round ends with the observation of a tail), the player \emph{loses} the game in that round; otherwise, the player \emph{wins} the game in that round.
Define a random variable $\wl(k,\ell)$ as the number of rounds that the player wins in the first $\ell$ flips.
\end{definition}

\begin{lemma}
\label{lem:upper_wl}
$\Pr(\wl(k,4ck \cdot 2^k) < 8ck) \ge 1-1/2^{ck}$
holds for any $k \ge 1$ and $c \ge 1$.
\end{lemma}
\begin{proof}
Let $\ell = 4 c k \cdot 2^k$.
The player wins the game with probability $2^{-k}$ at each round independently from other rounds. Therefore, by Chernoff bound (Lemma~\ref{lem:chernoff}), the player wins the game at least $8ck$ times in the first $\ell$ rounds with probability at most $\exp(-(4ck)/3) \le 2^{-ck}$.
Obviously, the player plays at most $\ell$ rounds
in the first $\ell$ flips, which completes the proof.
\end{proof}

\begin{lemma}
\label{lem:lower_wl} 
$\Pr(\wl(k,64ck \cdot 2^k) > 8ck) \ge 1-1/2^{ck}$ holds for any $k \ge 1$ and $c \ge 1$.
\end{lemma}
\begin{proof}
Let $\ell = 64 c k \cdot 2^k$.
The player wins the lottery game in at least $8ck$ rounds in the first $\ell$ flips if 
(i) the player plays at least $\ell/4$ rounds in the first $\ell$ flips,
and (ii) the player wins the game more than $8ck$ rounds in the first $\ell/4$ rounds. 
By Chernoff bound (Lemma~\ref{lem:chernoff}), 
event (i) occurs
with probability at least $1-\exp(-(32ck \cdot 2^k)/8) \ge 1-e^{-2ck}$.
The player wins the game with probability $2^{-k}$ at each round independently from other rounds. Therefore, by Chernoff bound (Lemma~\ref{lem:chernoff}), event (ii) occurs with probability at least $1-\exp(-(16ck)/8) \ge 1- e^{-2ck}$.
Thus, the lemma follows from the union bound since $2e^{-2ck} \le 2^{-2ck+1} \le 2^{-ck}$.
\end{proof}

We say that a signal at agent $u_i$ \emph{encounters}
an interaction when $u_i$ has an interaction.

\begin{proof}[Proof of Lemma~\ref{lem:keep_construction}]
We prove that the claim of the lemma holds when we define 
$\cmid$ as the set of all configurations in $\cnz$ where 
$u_i.\ho \le \cmax/2$ and $u_i.\mode = \Construct$ for all $u_i \in V$.


First, consider the execution $\Xi_{\ppl}(C_0,\sch)=C_0,C_1,\dots$,
where $C_0 \in \cnz$. By the definition of $\cnz$, there is always at least one leader in this execution.
Therefore, some leader generates a new signal within $O(n \log n)$ steps with probability 
at least $1-(1-1/n)^{n \log n} = 1-O(1/n)$.
Without loss of generality, we assume that $u_0$ is such a leader.
The generated signal visits all agents and resets their clocks in the next $6 n^2$ steps if all of the following events occur:
\begin{description}
    \item[(A)] $\seqr(0,n)$ occurs in the $6 n^2$ steps, 
    \item[(B)] The signal encounters at most $24 n$ interactions in the $6 n^2$ steps, and
    \item[(C)] The signal does not decrease its TTL to zero
in the first $24 n$ interactions it encounters.
\end{description}
Event (A) occurs w.h.p.\ by Lemma~\ref{lem:base}, 
while event (B) also occurs w.h.p.\ by Chernoff bound (Lemma~\ref{lem:chernoff}) because the signal encounters an interaction 
with probability $2/n$ at each step.
Event (C) holds w.h.p.\ because, for sufficiently large $\cmax=\Theta(\psi)$,
\begin{align*}
\Pr(\wl(\psi,24n) < \cmax) &\ge \Pr(\wl(\psi,24 \psi \cdot 2^\psi) < 48\psi)\\
&\ge 1 - 2^{-6\psi}\\
&\ge 1 - n^{-6},
\end{align*}
where we use Lemma~\ref{lem:upper_wl} with $k=\psi$ and $c = 6$ for the second inequality.
Thus, by the union bound, the clock of every agent is
reset to zero at least once in the $6 n^2$ steps.
Again, by Lemma~\ref{lem:upper_wl},
it holds w.h.p.\ that no agent increases its clock $\cmax/2$ times in the $6 n^2$ steps. 
Hence, the population will enter $\cmid$ within $O(n^2)$ steps w.h.p.

Next, we consider execution $\Xi_{\ppl}(C_0,\sch)=C_0,C_1,\dots$,
where $C_0 \in \cmid$.
In this execution, all agents remain in the construction mode until some agent increases its clock $\cmax/2$ times.
By Lemma~\ref{lem:upper_wl} with $k = \psi$ and $c=\cmax/(16\psi)$,
each agent $u_i$ increases its clock $\cmax/2$ or more times in its next $(\cmax/4) \cdot 2^\psi$ interactions with probability at most $2^{-\cmax/16}=O(1/n^2)$. By Chernoff bound (Lemma~\ref{lem:chernoff}), $u_i$ requires at least $(n \cmax / 16 ) \cdot 2^\psi$ steps to have $(\cmax/4) \cdot 2^\psi$ interactions with probability $1-O(1/n^2)$.
Hence, by the union bound, in an execution starting from a configuration in $\cmid$, all agents remain in the construction mode in the first $(n\cmax/16) \cdot 2^\psi = \Omega(n^2 \cmax)$ steps w.h.p.
\end{proof}

\begin{lemma}
\label{lem:absorbed}
Every resetting signal will be absorbed or disappear
in $O(n^2 \cmax)$ steps with probability $1-O(1/n^2)$.
\end{lemma}
\begin{proof}
The TTL of a resetting signal never increases unless it is absorbed and merged with another signal.
Hence, by Lemma~\ref{lem:lower_wl} with $k = \psi$ and $c = \cmax/(8\psi)$, the resetting signal is absorbed or disappears in the first $8\cmax\cdot 2^\psi$ interactions it encounters
with probability $1-2^{-\cmax/8} = 1-O(1/n^{2})$.
By Chernoff bound (Lemma~\ref{lem:chernoff}), a signal encounters at least $8\cmax \cdot 2^\psi$ interactions or disappears within $8n\cmax \cdot 2^\psi = O(n^2 \cmax)$ steps with probability $1-O(1/n^2)$.
Thus, the union bound gives the lemma.
\end{proof}

\begin{proof}[Proof of Lemma~\ref{lem:enter_detection}]
It is clear that a configuration $C$ is in $\cdet$ if 
there is no leader, all agents are in the detection mode,
and no resetting signal exists in $C$.
We will prove that $\Xi_{\ppl}(C_0,\sch)$ reaches such a configuration within $O(n^2 \log n)$ steps w.h.p. 
Assume $C_0 \in \lzero$. 
The number of resetting signals is monotonically non-increasing as long as there is no leader.
Thus, the union bound and Lemma~\ref{lem:absorbed} yield that the population reaches a configuration where 
there is no signal or there is a leader
within $O(n^2 \log n)$ steps w.h.p.
If there is no leader at this time,
by Lemma~\ref{lem:lower_wl}
with $k = \psi$ and $c = \cmax/(8\psi)$,
each agent $u_i$ increases its clock to $\cmax$
in its next $8\cmax \cdot 2^\psi$ interactions
or 
some agent becomes a leader during the period
with probability $1-O(1/n^2)$.
By Chernoff bound (Lemma~\ref{lem:chernoff}), 
$u_i$ has at least $8\cmax \cdot 2^\psi$ interactions within $8n\cmax \cdot 2^\psi = O(n^2 \cmax)$ steps with probability $1-O(1/n^2)$. Thus, the union bound gives the lemma.
\end{proof}

\subsection{Leader elimination}
\label{sec:elim}
We use the leader elimination part of the protocol given by Yokota, Sudo, and Masuzawa \cite{YSM21}
as Function $\elim()$ without any modification.
The pseudocode is given in Algorithm~\ref{al:le}.
In this module, leaders try to kill each other by firing \emph{bullets}
to decrease the number of leaders to one.
This module uses \emph{shields}, which removes bullets,
to avoid killing all leaders from the population.
An SS-LE protocol with bullets and shields in the ring
was first introduced in Fischer and Jiang \cite{FJ06}, but it requires $\Theta(n^3)$ expected steps for convergence.
Yokota \etal~\cite{YSM21} improve the bullets-and-shields-war strategy
to elect exactly one leader within $O(n^2)$ steps in expectation
and $O(n^2 \log n)$ steps with high probability.

The basic strategy is as follows.
A leader fires a bullet to kill another leader.
The fired bullet moves from left to right in the ring until it reaches a leader.
A leader does not always fire a bullet:
it fires a bullet only after it detects that the last bullet it fired
reaches a (possibly different) leader.
A leader may have a shield: a shielded leader is never killed even if it meets a bullet.
There are two kinds of bullets:
\emph{live bullets} and \emph{dummy bullets}.
A live bullet kills an unshielded leader.
However, a dummy bullet does not have the capability to kill a leader. 
When a leader decides to fire a new bullet,
the bullet becomes live or dummy with probability $1/2$ each.
When a leader fires a live bullet, 
it simultaneously generates a shield (if it is unshielded). 
When a leader fires a dummy bullet,
it breaks the shield (if it is shielded). 
Thus, roughly speaking, each leader is shielded with probability $1/2$ at a given time step.
Therefore, when a live bullet reaches a leader,
the leader becomes a follower with probability $1/2$.
This strategy is well designed: not all leaders kill each other
simultaneously because a leader must be shielded if
it fired a live bullet in the last shot. 
As a result, the number of leaders eventually decreases to one, but not to zero.

In what follows, we explain how we implement this strategy.
Each agent $v$ maintains variables $v.\bull \in \{0,1,2\}$, $v.\shield \in \{0,1\}$, and $v.\signalb \in \{0,1\}$.
As their names imply, $v.\bull=0$ (resp.~$v.\bull=1$, $v.\bull=2$)
indicates that $v$ is now conveying no bullet (resp.~a dummy bullet, a live bullet), while $v.\shield=1$ indicates that $v$ is shielded.
A variable $\signalb$ is used by a leader to detect
that the last bullet it fired already disappeared.
Specifically, $v.\signalb=1$ indicates that 
$v$ is propagating a \emph{bullet-absence signal}.
A leader always generates a bullet-absence signal
in its left neighbor when it interacts with its
left neighbor (Line~62).
This signal propagates from right to left (Line~62),
while a bullet moves from left to right (Lines~59 and 61).
A bullet disables a bullet-absence signal
regardless of whether it is live or dummy,
\ie $u_{i+1}.\signalb$ is reset to $0$
when two agents $u_i$ and $u_{i+1}$ such that $u_i.\bull > 0$
and $u_{i+1}.\signalb=1$ have an interaction (Line~60).
Thus, a bullet-absence signal propagates to a leader only after the last bullet fired by the leader disappears. 
When a leader $u_i$ receives a bullet-absence signal from its right neighbor $u_{i+1}$, $u_i$ waits for its next interaction
to extract randomness from the uniformly random scheduler.
At the next interaction, by the definition of the uniformly random scheduler, $u_{i}$ interacts with its right neighbor $u_{i+1}$ with probability $1/2$ and its left neighbor $u_{i-1}$ with probability $1/2$. In the former case, $u_i$ fires a live bullet and becomes shielded (Lines~51--52). In the latter case, $u_i$ fires a dummy bullet
and becomes unshielded (Lines~53-54).
In both cases, the received signal is deleted (Lines~52 and 54).
The fired bullet moves from left to right each time
the agent with the bullet, say $u_i$, interacts
with its right neighbor $u_{i+1}$ (Lines~59 and 61).
However, the bullet disappears without moving to $u_{i+1}$
if $u_{i+1}$ already has another bullet at this time.
Suppose that the bullet now reaches a leader.
If the bullet is live and the leader is not shielded
at that time, the leader is killed by the bullet (Line~57).
The bullet disappears at this time regardless of whether the bullet is live and/or the leader is shielded (Line~61). 
Lastly, whenever a new leader is created, it immediately fires a live bullet and becomes shielded (Lines~6 and 20).

Yokota \etal~\cite{YSM21} prove that this method decreases the number of
leaders to one within $O(n^2)$ steps in expectation (and thus within
$O(n^2 \log n)$ steps w.h.p.\,) starting from any configuration in
$\cnz \subseteq \lexist$, as long as no new leader is created during
this period. We formalize this fact as Lemma~\ref{lem:lone} in
Section~\ref{sec:convergence}.

\setcounter{AlgoLine}{50}
\begin{algorithm}[t]
\caption{$\elim()$\ \ \ $l$ is the initiator and $r$ is the responder of an interaction.}
\label{al:le}
\If{$l.\leader = l.\signalb = 1$}{
$(l.\bull,l.\shield,l.\signalb) \gets (2,1,0)$
}
\If{$r.\leader = r.\signalb = 1$}{
$(r.\bull,r.\shield,r.\signalb) \gets (1,0,0)$
}
\If{$l.\bull > 0$}{
\uIf{$r.\leader = 1$}{
$r.\leader \gets \begin{cases}
		     0 & l.\bull = 2 \land r.\shield = 0 \\
		     1 & \text{otherwise}
		    \end{cases}$\;
}
\Else{
$r.\bull \gets \begin{cases}
		      l.\bull & r.\bull = 0\\		       
		      r.\bull & \text{otherwise}
		      \end{cases}$\;

$r.\signalb \gets 0$\;
}
$l.\bull \gets 0$\;
}
$l.\signalb \gets \max(l.\signalb,r.\signalb,r.\leader)$
\end{algorithm}

\section{Correctness and convergence time}
\label{sec:correctness}
In this section, we prove Theorem~\ref{theorem:main} by showing that $\ppl$ is an SS-LE protocol on directed rings of any size $n$ given an integer $\psi$ satisfying $\log n \le \psi \le \log n  + O(1)$
and that the convergence time of $\ppl$ is $O(n^2 \log n)$ steps w.h.p.
First, in Section~\ref{sec:safe}, we introduce a set $\srl$ of configurations and prove that every configuration in $\srl$ is safe.
Next, in Section~\ref{sec:convergence}, we prove that the population reaches a configuration in $\srl$ within $O(n^2 \log n)$ steps w.h.p.\ starting from any configuration. 

We use several functions whose return values
depend on a configuration,
such as $\distll(i)$ and $\distrl(i)$.
When a configuration should be specified,
we explicitly write a configuration as the first argument
of those functions.
For example, 
we denote $\distll(C,i)$ and $\distrl(C,i)$
to denote $\distll(i)$ and $\distrl(i)$
in a configuration $C$, respectively.

\subsection{Safe configurations}
\label{sec:safe}

We introduce a set of safe configurations $\srl$ here.
Define $\lone$ as the set of configurations where 
there is exactly one leader. 
In the rest of this paper, 
whenever we consider a configuration $C \in \lone$, we always assume that $u_0$ is the unique leader in $C$ without loss of generality.

Of course, $\srl \subseteq \lone$ must hold.
Moreover, an execution starting from $\srl$ must keep exactly one leader. 
Thus, it must not kill the unique leader $u_0$.
Let us discuss what configurations have this property.

We say that a live bullet located at agent $u_i$ is \emph{peaceful} when the following predicate holds:
\begin{align*}
\peaceful(i) &\predef
\left(
\begin{aligned}
&(\distll(i) \neq \infty \to u_{i-\distll(i)}.\shield = 1) \\
& \land \forall j \in [0,\distll(i)]: u_{i-j}.\signalb = 0
\end{aligned}
\right).
\end{align*}
Note that when $\distll(i)=\infty$, $\peaceful(i)$ requires that no bullet-absence signal exists anywhere in the population.
Peaceful bullets have the following useful properties.

\begin{lemma}
\label{lem:never_kill}
If there exists exactly one leader in the population, then a peaceful bullet never kills the leader.
\end{lemma}
\begin{proof}
By the definition of a peaceful bullet, the unique leader is shielded and hence cannot be killed.
\end{proof}

\begin{lemma}
\label{lem:new_live_bullet}
Every newly fired live bullet is peaceful. 
\end{lemma}
\begin{proof}
When a leader fires a live bullet, it becomes shielded and disables any bullet-absence signal.
\end{proof}

\begin{lemma}
\label{lem:once_peaceful}
A peaceful bullet never becomes non-peaceful.
\end{lemma}

\begin{proof}
Assume for contradiction that a configuration $C$ changes to $C'$ via an interaction
$e_j = (u_j,u_{j+1})$, and a peaceful bullet located at an agent $u_i$ in $C$
becomes non-peaceful in $C'$.

First, consider the case $j = i$, where the bullet moves to the right.
If $u_{i+1}$ is a leader in $C$ or a new leader is created at $u_{i+1}$ by this interaction,
the bullet disappears and thus does not exist in $C'$, a contradiction.
Otherwise, the bullet moves to $u_{i+1}$, disabling the signal at $u_{i+1}$ (if any).
No new signal is generated and no leader is killed.
Thus, the bullet remains peaceful in $C'$, a contradiction.

Next, consider the case $j \neq i$ and $C \in \lexist$.
The nearest left leader $u_{i-\distll(C,i)}$ is not killed
because it is shielded in $C$,
and it remains shielded since $u_{i-\distll(C,i)}.\signalb = 0$ in $C$.
A new leader may be created at $u_{j+1}$, but it is shielded at creation (Lines~6 and 20).
No new bullet-absence signal is generated in the agents
$u_{i-\distll(C',i)}, \dots, u_i$
because $u_{i-\distll(C',i)+1}, \dots, u_i$
are all followers (note that $\distll(C',i) \le \distll(C,i)$).
Therefore, the bullet remains peaceful in $C'$, a contradiction.

Finally, consider the case $j \neq i$ and $C \in \lzero$.
In this case, by the definition of a peaceful bullet, there is no
bullet-absence signal in the population in $C$.
Even if a new leader is created at $u_{j+1}$,
a bullet-absence signal is created only at $u_j$,
which does not lie in the interval $u_{j+1}, u_{j+2}, \dots, u_i$.
Hence, the bullet remains peaceful in $C'$, a contradiction.
\end{proof}

Initially, there may be one or more non-peaceful live bullets. 
However, by Lemmas~\ref{lem:never_kill},~\ref{lem:new_live_bullet}, and~\ref{lem:once_peaceful},
once the population reaches a configuration where
every live bullet is peaceful and there are one or more leaders,
the number of leaders never becomes zero.
Formally, we define the set of such configurations
as follows:
\begin{align*}
 \cpb = \left \{ C \in \lexist \longmid \forall u_i \in V: C(u_i).\bull = 2 \Rightarrow \peaceful(C,i) \right \}
\end{align*}
\noindent
The following lemma immediately follows from Lemmas~\ref{lem:never_kill},~\ref{lem:new_live_bullet}, and~\ref{lem:once_peaceful}.
\begin{lemma}
\label{lem:cpb_closed} 
$\cpb$ is closed. 
\end{lemma}
\begin{corollary}
\label{cor:cpb_cnz}
$\cpb \subseteq \cnz$.
\end{corollary}

Is $\cpb \cap \lone$ closed?
If the answer is yes,
every configuration in $\cpb \cap \lone$ is safe;
thus we can define $\srl = \cpb \cap \lone$.
Unfortunately, the answer is no:
an execution starting from a configuration in $\cpb \cap \lone$ 
never kills the unique leader $u_0$, however 
it may create a new leader.
We must specify the condition that
the population never creates a leader.

First, we define $\cdl \subseteq \cpb \cap \lone$ as the set of
configurations in which $\distb$ and $\last$ are correctly computed at
every agent. Formally, a configuration $C \in \cpb \cap \lone$ belongs
to $\cdl$ if, for every $i \in [0,n-1]$,
\[
u_i.\distb = i \bmod 2\psi
\quad\land\quad
\bigl(u_i.\last = 1
\Leftrightarrow
i \in [\psi(\zeta-1),n-1]\bigr).
\]
(Recall $\zeta = \lceil n /\psi \rceil$.)

Second, we consider the condition on a variable $\id$.
We define 
\[S_i = u_{i\psi},u_{i\psi+1},\dots,u_{i\psi+\psi-1}\]
for any $i \in [0,\zeta - 2]$
and call $S_i$ the $i$-th segment.
Each segment $S_i$ is a sequence of agents,
but we sometimes use $S_i$ as a set by abuse of notation.
If a configuration $C$ is safe, 
we expect $\iota(C,S_{i+1}) = \iota(C,S_i) +1 \bmod 2^\psi$
for any $i \in [0,\zeta - 3]$,
that is, $C$ must be perfect. 
This condition is still insufficient because a token may hold an inconsistent value so that it will change variable $\id$ or create a new leader even if the current configuration is perfect. 
Thus, we also expect that there is no such token. 
For any token $z$ located at $u_k$,
we call agent $u_{k + u_k.\token[1]}$ the \emph{target} of $z$,
denoted by $T(z)$.
Recall that a token changes its moving direction every time it reaches its target.
For any $i \in [0,\zeta - 1]$,
we say that a valid token $z$ at $u_k$ is 
\emph{working for} $(S_i,S_{(i+1 \bmod{\zeta})})$ if 
\[T(z) \in S_{(i+1 \bmod{\zeta})} \land u_k.\token[1] >0, \quad \text{or} \quad
T(z) \in S_{i} \land u_k.\token[1] < 0.
\]

Consider any valid token $z$ working for $(S_i, S_{i+1})$ with $i \in [0,\zeta-3]$.
Since $z$ is valid (See Definition~\ref{def:invalid}), $i$ is an even number if and only if $z$ is black. For any $x \in [0,\psi-1]$, we say that $z$ is
in the $x$-th round if its target is $u_{i\psi + \psi + x}$ or $u_{i\psi+x+1}$.
In each round $x$, the token $z$ moves right from $u_{i\psi + x}$, reaches the target $u_{i\psi + \psi + x}$, changes the direction to the left, and finally reaches the target $u_{i\psi + x +1}$.
Then, we can define the correctness of tokens.
Recall that $\token[2]$ stores the binary value to set (check) at the target and $\token[3]$ stores the carry flag.
\begin{definition}
\label{def:correct}
Let $i \in [0,\zeta-3]$.
Consider that a valid token $z$ working for $(S_i,S_{i+1})$ is in the $x$-th round and is located at $u_k$.
Let $j$ be the minimum integer in $[0,\psi-1]$ such that 
$u_{i\psi + j }.\id = 0$. Define $j = \psi$ if no such $j$ exists in $[0,\psi-1]$.
Then, we say that $z$ is \emph{correct}
if the following conditions hold:
\[
u_k.\token[2] = u_{i\psi + x}.\id \xor \mathbf{1}_{x \le j},
\qquad 
u_k.\token[3] = \mathbf{1}_{x < j}.
\]
\end{definition}
\noindent 
The following remarks immediately follow from the definition
of correct tokens.
\begin{remark}
\label{rem:correct}
Let $i \in [0,\zeta - 3]$,
and let $z$ be a correct token
which is working for $(S_i,S_{i+1})$,
located at $u_k$, and in the $x$-th round in a configuration $C \in \cdl$.
Then, $z$ satisfies $u_k.\token[2] = b_x$, where $b_0,b_1,\dots,b_{\psi-1}$ is a sequence of bits such that $\sum_{j \in [0,\psi-1]} b_j \cdot 2^{j} = \iota(S_i) + 1 \bmod 2^\psi$ in $C$. 
\end{remark}

\begin{remark}
\label{rem:get_incorrect}
A correct token working for $(S_i,S_{i+1})$ with $i \in [0,\zeta-3]$ becomes incorrect only when $\iota(S_i)$ changes, as long as the population is in $\cdl$.
\end{remark}

We do not define the correctness of tokens working for
$(S_{\zeta-2},S_{\zeta-1})$ or $(S_{\zeta-1},S_0)$.
A token can take a harmful action only when it reaches its target while
moving right.
In $\cdl$, every agent in the last segment $S_{\zeta-1}$ has
$\last=1$, so no token working for $(S_{\zeta-2},S_{\zeta-1})$ can
reach its target while moving right (i.e., while $\token[1]>0$)
because of Line~13.
Thus, we can safely ignore such tokens.
On the other hand, a right-moving token working for
$(S_{\zeta-1},S_0)$ may be harmful if it can reach its target in $S_0$
without passing through the last segment, where token movement is blocked.
We explicitly exclude such tokens from configurations in $\srl$.
\begin{definition}
We say that a token $z$ located at $u_k$ is \emph{risky} if
\[
u_k.\token[1]>0
\quad\text{and}\quad
u_{k+h} \in S_0
\text{ for every } h \in [0,u_k.\token[1]].
\]
\end{definition}

Now, we have all the ingredients to define $\srl$.
\begin{definition}
\label{def:srl}
Define $\srl$ as the set of configurations in $\cdl$ in which
\begin{itemize}
\item there are no risky tokens, and
\item for every $i \in [0,\zeta-3]$, all valid tokens working for
$(S_i,S_{i+1})$ are correct and
\[
\iota(S_{i+1}) = \iota(S_i) + 1 \bmod 2^\psi.
\]
\end{itemize}
\end{definition}

\begin{lemma}
\label{lem:safe}
$\srl$ is closed. Every configuration in $\srl$ is safe.
\end{lemma}
\begin{proof}
The unique leader is never killed in any transition from a configuration in $\srl \subseteq \cpb \cap \lone$.
Therefore, if $\srl$ is closed, every configuration in $\srl$ is safe.
Thus, it suffices to prove the first claim: $\srl$ is closed.
Let $C \in \srl, C' \in \call(\ppl)$ be configurations
such that $C$ changes to $C'$ by one interaction.
Since $C \in \srl \subseteq \cdl$, 
no agent becomes a new leader by observing inconsistency of $\distb$ (Line 6).
Moreover, since all valid tokens working for $(S_i,S_{i+1})$ with $i \in [0,\zeta-3]$ are correct
and no risky token exists in $C$, 
by Remark~\ref{rem:correct}, no agent changes variable $\id$
and no agent becomes a new leader in that interaction.
The population can deviate from $\cdl$ only when a new leader is created. 
Thus, $C' \in \cdl$ and 
$\iota(C',S_{i+1}) = \iota(C',S_i) + 1 \bmod{2^\psi}$ holds for all $i \in [0,\zeta-3]$.
By Remark~\ref{rem:get_incorrect}, all correct tokens that appear in $C$
are still correct in $C'$.
Moreover, 
whenever a token is generated by a border agent in $S_i$ with $i \in [0,\zeta-3]$, 
it is correct by definition of correctness.
Thus, all valid tokens working for $(S_i, S_{i+1})$ with $i \in [0,\zeta-3]$ are still correct in $C'$.
Since $C \in \cdl$, no token can move from $u_{n-1}$ to $u_0$ because
$u_{n-1}.\last=1$. Therefore, no new risky token arises in $C'$.
Hence, $C'$ is still in $\srl$, which proves the lemma.
\end{proof}

\subsection{Convergence}
\label{sec:convergence}
In this subsection, we prove
that an execution of $\ppl$ 
starting from any configuration in $\call(\ppl)$
reaches a configuration in $\srl$
within $O(n^2 \log n)$ steps w.h.p.
We give this upper bound in two steps:
we first show that execution $\Xi_{\ppl}(C_0,\sch)$
enters $\cpb$ within $O(n^2 \log n)$ steps w.h.p.\ for any $C_0 \in \call(\ppl)$, 
and next show that 
$\Xi_{\ppl}(C_0,\sch)$
enters $\srl$ within $O(n^2 \log n)$ steps w.h.p.\ if $C_0 \in \cpb$.

\begin{lemma}
\label{lem:get_peaceful}
Every live bullet disappears or becomes peaceful before $(\seqr(0,n))^2$ completes.
\end{lemma}
\begin{proof}
Assume for contradiction that a non-peaceful live bullet $b$, which is initially at $u_i$, 
remains alive (i.e., does not disappear by meeting a leader) and non-peaceful when $\seqr(i,n)$ completes.
Whenever $b$ moves from $u_{i+j}$ to $u_{i+j+1}$, it disables a bullet-absence signal at $u_{i+j+1}$ (Line~60).
Thus, $b$ moves exactly $n$ times by the time $\seqr(i,n)$ completes, without meeting a leader.
Let $X$ be the number of times that $b$ moves to the right; then $b$ is located at $u_{i+X}$.

No leader is created at any agent among $u_i, u_{i+1}, \dots, u_{i+X}$.
Otherwise, suppose that such an event first occurs at $u_{i+j}$, where $0 < j \le X$.
Then $b$ becomes peaceful because the new leader $u_{i+j}$ is shielded,
and no bullet-absence signal exists at $u_{i+j}, u_{i+j+1}, \dots, u_{i+X}$, a contradiction.

Therefore, when $\seqr(i,n)$ completes, there is no leader and no bullet-absence signal in the population,
implying that the bullet is peaceful, again a contradiction.
\end{proof}

\begin{lemma}
\label{lem:cpb}
Let $C_0 \in \call(\ppl)$. 
Then, $\Xi = \Xi_{\ppl}(C_0,\sch)$ enters $\cpb$ within $O(n^2 \log n)$ steps
 w.h.p.
\end{lemma}

\begin{proof}
By Lemmas~\ref{lem:new_live_bullet},~\ref{lem:once_peaceful}, and~\ref{lem:get_peaceful},
the population reaches a configuration without any non-peaceful bullet before $(\seqr(0,n))^2$ completes,
after which no non-peaceful bullet ever appears.
By Lemma~\ref{lem:base}, this requires $O(n^2)$ steps w.h.p.

Thus, it suffices to show that a leader is created within $O(n^2 \log n)$ steps w.h.p.
By Lemma~\ref{lem:enter_detection}, the population reaches, within
$O(n^2 \log n)$ steps w.h.p., a configuration from which all agents
remain in the detection mode until a new leader is created.
We may assume that $n$ is divisible by $2\psi$ and that $u_{i+1}.\distb = u_i.\distb + 1 \bmod 2\psi$ holds for all $i \in [0,n-1]$,
because otherwise the inconsistency is detected and a new leader is created (Line~6) within $O(n \log n)$ steps w.h.p. We assume $u_0.\distb = 0$ without loss of generality.
Then, all agents have $\last = 0$ before or when $\seql(0,n)$ completes, which requires $O(n^2)$ steps w.h.p.\ by Lemma~\ref{lem:base}. Thereafter, no agent changes $\distb$ or $\last$ until a leader is created. 
Let $S_i = u_{i \psi}, u_{i \psi + 1},\dots,u_{i \psi + \psi - 1}$
for all $i \in [0,\zeta-1]$, where $\zeta = n/\psi$.
At this point, by Lemma~\ref{lem:impossible}, 
there must exist $i \in [0,\zeta-1]$ such that 
$\iota(S_{i+1}) \neq \iota(S_i) + 1 \bmod 2^\psi$, 
where the addition over the segment indices is modulo $\zeta$.
Agents in the detection mode never update variable $\id$, thus 
$\iota(S_i)$ never changes until a leader is created. 
Since a newly generated token is always correct,
by Lemma~\ref{lem:complete}, the inconsistency $\iota(S_{i+1}) \neq \iota(S_i) + 1 \bmod 2^\psi$ will be detected and a new leader will be created before or when $(\seqr(i\psi,2\psi-1)\cdot \seql(i\psi+2\psi-1,2\psi-1))^{2\psi}$ completes, which requires $O(n \log^2 n)$ steps w.h.p.\ by Lemma~\ref{lem:base}.
To conclude, a new leader is created within $O(n^2 \log n)$ steps w.h.p.
\end{proof}

Next, we prove the second claim, \ie that
$\Xi_{\ppl}(C_0,\sch)$ enters $\srl$ within $O(n^2 \log n)$ steps
w.h.p.\ for any $C_0 \in \cpb$.
For the analysis, we introduce an auxiliary protocol $\ppl'$.
Protocol $\ppl'$ is identical to $\ppl$, except that it always executes
the branch for $\mode=\Construct$ in commands whose behavior depends on
$\mode$, regardless of the actual value of $\mode$.
Thus, in Algorithms~\ref{al:create} and~\ref{al:token}, $\ppl'$ updates
$\distb$ and $\id$ as in the construction mode instead of creating a new
leader upon detecting an inconsistency.
All other commands, including $\determ()$ and $\elim()$, are executed
exactly as in $\ppl$.
In particular, $\ppl$ and $\ppl'$ behave identically whenever all agents
are in the construction mode.

\begin{lemma}[\cite{YSM21}]
\label{lem:lone}
Let $D_0 \in \cpb$. 
Then, $\Xi' = \Xi_{\ppl'}(D_0,\sch)$ enters $\cpb \cap \lone$ within $O(n^2 \log n)$ steps w.h.p.
\end{lemma}
\begin{proof}
Yokota \etal~\cite{YSM21} shows that
$\elim()$ decreases the number of leaders to one 
within $O(n^2)$ steps in expectation
as long as a new leader is not created.
Since $\ppl'$ does not have the ability to create a new leader,
$\Xi'$ reaches a configuration in $\cpb \cap \lone$
within $O(n^2)$ steps in expectation.
By Markov's inequality, $\Xi'$ enters $\cpb \cap \lone$
within $O(n^2)$ steps with probability $1/2$.
Since $\cpb$ is closed by Lemma~\ref{lem:cpb_closed}, 
$\Xi'$ enters $\cpb \cap \lone$ within $O(n^2 \log n)$ steps
with probability at least $1-(1-1/2)^{\log n}=1-O(1/n)$.
\end{proof}

\begin{lemma}
\label{lem:srl_variant}
Let $D_0 \in \cpb$. 
Then, $\Xi' = \Xi_{\ppl'}(D_0,\sch)$ enters $\srl$ within $O(n^2 \log n)$ steps w.h.p.
\end{lemma}
\begin{proof}
We can assume $D_0 \in \cpb \cap \lone$ by Lemma~\ref{lem:lone}.
As mentioned above, we always assume that $u_0$ is the unique leader in a configuration in $\lone$.
Then, $e_{n-1}\cdot\seqr(0,n-1)\cdot \seql(0,n)$ lets the population
enter $\cdl$, which requires $O(n^2)$ steps w.h.p.\ by Lemma~\ref{lem:base}.
Indeed, the first interaction $e_{n-1}$ sets $u_0.\distb=0$ because
$u_0$ is the unique leader.
Thereafter, $\seqr(0,n-1)$ propagates the correct distance values
clockwise from $u_0$.
Finally, $\seql(0,n)$ propagates the information on the last segment
counter-clockwise, so that $u_i.\last=1$ holds exactly for the agents
in the last segment.
Moreover, during the next $\seqr(0,\psi)^2$, every risky token becomes
non-risky, and no risky token arises thereafter.
Since a newly generated token is always correct,
the next $(\seqr(0,2\psi-1)\cdot \seql(2\psi-1,2\psi-1))^{2\psi} \cdot (\seqr(\psi,2\psi-1)\cdot \seql(3\psi-1,2\psi-1))^{2\psi} \cdot \dots \cdot (\seqr((\zeta-3)\psi,2\psi-1)\cdot \seql((\zeta-1)\psi-1,2\psi-1))^{2\psi}$, where $\zeta= \lceil n/\psi \rceil$,
lets the population enter $\srl$
by Lemma~\ref{lem:complete} and Remark~\ref{rem:get_incorrect}. This requires $O(n\psi^2 \zeta) = O(n^2\psi) = O(n^2 \log n)$ steps w.h.p.\ by Lemma~\ref{lem:base}.
\end{proof}

\begin{lemma}
 \label{lem:srl}
Let $C_0 \in \cpb$. 
Then, $\Xi = \Xi_{\ppl}(C_0,\sch)$ enters $\srl$ within $O(n^2 \log n)$ steps w.h.p.\ for a sufficiently large $\cmax = \Theta(\psi) = \Theta(\log n)$.
\end{lemma}
\begin{proof}
By Lemma~\ref{lem:keep_construction},
there exists a set $\cmid$ of configurations such that
(i) $\Xi$ enters $\cmid$ within $O(n^2)$ steps
 w.h.p., 
and (ii) letting $C_t \in \cmid$, all agents are in the construction mode in the first $\Omega(\cmax n^2)$ steps of $\Xi_{\ppl}(C_t,\sch) = C_t, C_{t+1},C_{t+2},\dots$ w.h.p.
Thus, $\Xi_{\ppl}(C_t,\sch)$ does not create a leader in the first $\Omega(\cmax n^2)$ steps w.h.p.
Therefore, $\Xi_{\ppl}(C_t,\sch)$ and $\Xi_{\ppl'}(C_t,\sch)$
are the same in the first $\Omega(\cmax n^2)$ steps w.h.p. Since $\cmax$ is a sufficiently large $O(\log n)$ value,
by Lemma~\ref{lem:srl_variant}, we conclude that
$\Xi$ enters $\srl$ within $O(n^2 \log n)$ steps w.h.p.
\end{proof}

Theorem~\ref{theorem:main} immediately follows from Lemmas~\ref{lem:safe},~\ref{lem:cpb}, and~\ref{lem:srl}.

\begin{algorithm}[t]
\caption{$\por$\ \ \   $a$ is the initiator and $b$ is the responder of an interaction.}
\label{al:ro}

\VarAgent{
\\
\varspace $\iro, \cone, \ctwo \in [0,\colnum-1]$,   \hfill \mycommfont{// input variables}\\ 
\varspace $\dir \in [0,\colnum-1]$,
\varspace $\strong \in \{0,1\}$
}

\For{$v \in \{a,b\}$}{
\If{$v.\dir \notin \{v.\cone,v.\ctwo\}$}{
$v.\dir \gets v.\cone$\;
}
}
\uIf(\tcp*[f]{two heads meet}){$a.\dir = b.\iro \land b.\dir = a.\iro$}{
\uIf{$a.\strong = 0 \land b.\strong = 1$}{
$a.\dir \gets \min \{c \in \{a.\cone, a.\ctwo\} \mid c \neq b.\iro\}$ \tcp*{flip $a.\dir$}
$(a.\strong,b.\strong) \gets (1,0)$\;
}
\Else{
$b.\dir \gets \min \{c \in \{b.\cone,b.\ctwo\} \mid c \neq a.\iro\}$ \tcp*{flip $b.\dir$}
$(a.\strong,b.\strong) \gets (0,1)$\;
}
}
\uElseIf{$a.\dir = b.\iro$}{
$a.\strong \gets 0$\;
}
\ElseIf{$b.\dir = a.\iro$}{
$b.\strong \gets 0$\;
}
\end{algorithm}

\section{Removal of the orientation assumption}
\label{sec:orient}
In this section, we show how to remove the assumption that the population is a \emph{directed} ring.
Specifically, we present a self-stabilizing ring orientation protocol that uses only $O(1)$ states
and converges in $O(n^2 \log n)$ steps w.h.p.\ on any undirected ring $G=(V,E)$. 
As in the previous sections, we assume $V=\{u_0,u_1,\dots,u_{n-1}\}$, but do not assume
$E= \{(u_i,u_{i+1}) \mid i \in [0,n-1]\}$.
Instead, we assume
\[E= \{(u_i,u_{i+1}) \mid i \in [0,n-1]\} \cup \{(u_{i+1},u_i) \mid i \in [0,n-1]\}\]
and
\[\Pr(\Gamma_t = (u_i,u_{i+1})) = \Pr(\Gamma_t = (u_{i+1},u_{i})) = \frac{1}{2n}\]
for every $t \ge 0$ and $i \in [0,n-1]$ in this section. 

We define a self-stabilizing ring orientation protocol as follows.
\begin{definition}[Self-stabilizing Ring Orientation (SS-RO)]
\label{def:ro}
Let $P$ be a protocol that has output variables 
$\iro, \dir \in [0,\colnum-1]$,
where $\colnum$ is a positive integer.
We say that a configuration $C$ of $P$ is \emph{safe}
if (i) $u_i.\iro \neq u_{i+2}.\iro$ for any $i \in [0,n-1]$,
(ii) $u_i.\dir = u_{i+1}.\iro$ holds for any $i \in [0,n-1]$
or $u_j.\dir = u_{j-1}.\iro$ holds for any $j \in [0,n-1]$,
and (iii)
at every configuration reachable from $C$,
no agent changes the values of the output variables 
$\iro$ or $\dir$.
A protocol $P$ is a \emph{self-stabilizing ring orientation (SS-RO) protocol}
if $\Xi_{P}(C_0,\sch)$ reaches a safe configuration
with probability $1$ for any configuration $C_0 \in \call(P)$.
\end{definition}

The first condition is the specification of a well-known problem called
\emph{two-hop coloring},
by which each agent can distinguish its two neighbors.
The second condition gives the agents a common sense of direction:
each agent points at one of its neighbors,
and two neighboring agents $u_i$ and $u_{i+1}$ never point at each other,
\ie $u_i.\dir = u_{i+1}.\iro \ \Leftrightarrow\ u_{i+1}.\dir \neq u_i.\iro$.

Fortunately, several self-stabilizing two-hop coloring protocols are
known~\cite{AAF+08,SOK+18,RSV26}.
The protocol of Sudo~\etal~\cite{SOK+18} uses $O(n^4)$ states and converges in
$O(mn^2 \log n)$ steps on arbitrary graphs.
With a slight modification, however, this protocol uses only a constant number of
states and converges in $O(n^2)$ steps in expectation, and thus in
$O(n^2 \log n)$ steps w.h.p. on rings.
For completeness, we describe such a modification in~\ref{sec:two-hop}.
Alternatively, we could directly use the more recent two-hop coloring
protocol by Rybicki, Solnerzik, and Vacus~\cite{RSV26}, which uses
$O(2^\Delta)$ states and converges in $O(\Delta m \log n)$ expected
steps, where $\Delta$ is the maximum degree.
In particular, $\Delta=2$ and $m=n$ on rings, so this protocol uses
$O(1)$ states and converges in $O(n\log n)$ expected steps.\footnote{
We describe the modified protocol of Sudo~\etal~\cite{SOK+18} in the appendix
rather than directly using the protocol of Rybicki~\etal~\cite{RSV26} because
the latter has not yet appeared in the proceedings of a peer-reviewed conference
or in a journal.
}

Moreover, once a two-hop coloring is achieved, 
the agents can store the set of colors of their two neighbors 
in $O(n \log n)$ steps w.h.p.\ using $O(1)$ states by the following simple protocol:
\begin{quote} 
Each agent memorizes the two distinct colors that it has observed most recently. 
\end{quote}

\begin{note}
We present an SS-RO protocol under the assumptions that the first condition of Definition~\ref{def:ro} is always satisfied,
each agent maintains two variables $\cone, \ctwo \in [0,\colnum-1]$,
and these variables always satisfy 
$\{u_i.\cone,u_i.\ctwo\} = \{u_{i-1}.\iro,u_{i+1}.\iro\}$
for every $i \in [0,n-1]$.
These assumptions do not lose generality.
We execute the two-hop coloring protocol (\ref{sec:two-hop})
and the above process for memorizing two colors in parallel with our SS-RO protocol.
Therefore, from some point, these assumptions are always satisfied.
\end{note}

The proposed SS-RO protocol, named $\por$, is shown in Algorithm~\ref{al:ro}.
In this pseudocode, $a$ and $b$ denote the initiator and responder,
respectively.
We use the condition
\[
u_i.\dir \in \{u_i.\cone,u_i.\ctwo\}
\]
for every agent $u_i$ as an invariant in our analysis.
This invariant may not hold initially because the initial states are
assigned adversarially.
However, whenever an agent $u_i$ participates in an interaction with
$u_i.\dir \notin \{u_i.\cone,u_i.\ctwo\}$, it sets $u_i.\dir$ to
$u_i.\cone$ (Lines~63--65).
Therefore, after every agent has participated in at least one interaction,
all agents satisfy the invariant.
This requires $O(n\log n)$ steps both in expectation and w.h.p.
In the following analysis, we ignore this initial part of the execution.

We call a maximal sequence of distinct agents
$u_i,u_{i+1},\dots,u_{i+k}$ a \emph{right segment} if every agent in the
sequence points to its right neighbor; that is,
\[
    u_{i+j}.\dir = u_{i+j+1}.\iro
    \qquad\text{for every } j \in [0,k].
\]
The last agent $u_{i+k}$ in the sequence is called the \emph{head} of the segment.
Similarly, we call a maximal sequence of distinct agents
$u_i,u_{i-1},\dots,u_{i-k}$ a \emph{left segment} if every agent in the
sequence points to its left neighbor; that is,
\[
    u_{i-j}.\dir = u_{i-j-1}.\iro
    \qquad\text{for every } j \in [0,k].
\]
The last agent $u_{i-k}$ in the sequence is called the \emph{head} of the segment.
We collectively refer to right and left segments simply as \emph{segments}.
Note that every agent belongs to exactly one segment, whose size ranges from
one to $n$.
Ring orientation is achieved when a segment of size $n$ occupies the entire ring.

In $\por$, each head always tries to extend its segment.
When two heads meet, only one of them \emph{wins} and extends its segment by flipping the loser's $\dir$ (Lines~66--72). 
No other rules update variable $\dir$. Therefore, the number of segments in the population is monotonically non-increasing.
The losing segment decreases its size by one, and disappears if its size becomes zero.
In this case, the number of segments decreases by two because the two neighboring segments merge.
The only exception is the case where there are only two segments and one of them disappears.
Then, the number of segments decreases from two to one, at which point the ring orientation is achieved.

To speed up the protocol, we introduce a variable $\strong \in \{0,1\}$.
An agent $u_i$ is called \emph{strong} if $u_i.\strong = 1$; otherwise, the agent is \emph{weak}. 
We use the following rule when two heads meet:
\begin{itemize}
    \item If exactly one of them is strong, the strong head wins. 
    \item If both are strong, or both are weak, the initiator wins.    
\end{itemize}
In any case, the loser flips its $\dir$ and becomes strong, while the winner becomes weak.
Therefore, when a segment extends by one, its new head becomes strong.
However, if a segment absorbs another segment and increases its length by more than one,
the new head of the resulting segment may be weak.

We call a strong non-head agent an \emph{obstacle}.
Whenever a strong agent $u$ interacts with an agent $v$
such that $u.\dir = v.\iro$ but $v.\dir \neq u.\iro$,
$u$ finds itself to be an obstacle and becomes weak (Lines~73--76).

\begin{lemma}
\label{lem:obstacle}
An agent becomes an obstacle only when it loses a head-to-head interaction
and consequently the segment to which it belongs disappears.
\end{lemma}

\begin{proof}
When an agent belonging to a segment of size at least two loses a head-to-head interaction,
it remains a head after the interaction and thus does not become an obstacle.
Therefore, it suffices to show that an agent becomes an obstacle only when it loses a head-to-head interaction.

Assume for contradiction that an agent $u_i$ becomes an obstacle by an interaction between $u_j$ and $u_{j+1}$,
but either $i \notin \{j,j+1\}$ or at least one of $u_j$ and $u_{j+1}$ is a non-head.

Suppose first that $i \notin \{j,j+1\}$.
Then, $u_i$ must be strong and a head before the interaction between $u_j$ and $u_{j+1}$,
and become a non-head by that interaction.
However, a head becomes a non-head only when it interacts with another head,
contradicting $i \notin \{j,j+1\}$.

Suppose next that at least one of $u_j$ and $u_{j+1}$ is a non-head.
Then, no agent updates variable $\dir$ or becomes strong.
Hence, no agent becomes an obstacle, a contradiction.
\end{proof}


\begin{theorem}
\label{theorem:ssro}
There exists a self-stabilizing ring orientation (SS-RO) protocol that converges in $O(n^2 \log n)$ steps
both w.h.p.\ and in expectation and uses $O(1)$ states per agent. 
\end{theorem}

\begin{proof}
We prove that protocol $\por$ solves SS-RO within $O(n^2 \log n)$ steps both w.h.p.\ and in expectation using $O(1)$ states.
Let $S_0,S_1,\dots,S_{k-1}$ be the segments in the initial configuration, where 
$S_i$ and $S_{i+1}$ are adjacent for every $i \in [0,k-1]$, and 
$S_{2j}$ and $S_{2j+1}$ face each other, i.e., their heads are neighboring and point each other by variable $\dir$, for every $j \in [0,k/2-1]$,
where segment indices are taken modulo $k$.
Note that $k$ must be even unless $k=1$, in which case the ring orientation has already been achieved.

We first prove the following claim. 

\begin{claim}
\label{claim:disappear}
If $k \ge 2$, then for every $j \in [0,k/2-1]$, at least one of 
$S_{2j+1}, S_{2j+2}, S_{2j+3}$, and $S_{2j+4}$
disappears within $O(n^2)$ steps with probability $1-O(1/n^3)$.
\end{claim}

\begin{proof}
By Lines~73--76, each obstacle (i.e., strong non-head agent) becomes weak with probability $1/n$ at each step. 
Moreover, as long as all four segments survive, Lemma~\ref{lem:obstacle} implies that
the number of obstacles in $S_{2j+2}$ and $S_{2j+3}$ is monotonically non-increasing.
Hence, unless one of $S_{2j+1},S_{2j+2},S_{2j+3}$, and $S_{2j+4}$ disappears earlier,
all obstacles in $S_{2j+2}$ and $S_{2j+3}$ disappear within $O(n \log n)$ steps
with probability $1-O(1/n^3)$.

Thereafter, once the head of $S_{2j+2}$ (resp.~$S_{2j+3}$) wins,
it keeps winning until either $S_{2j+3}$ or $S_{2j+4}$
(resp.~$S_{2j+2}$ or $S_{2j+1}$) disappears.
Hence, some of the four segments disappear while the interaction sequence
$\seqr'(0,n)^2\cdot \seql'(0,n)^2$ occurs, where
\begin{align*}
&\seqr'(0,n) = (u_0,u_1),(u_1,u_2),\dots,(u_{n-1},u_0),\\
&\seql'(0,n) = (u_0,u_{n-1}),(u_{n-1},u_{n-2}),\dots,(u_1,u_0).
\end{align*}
This requires $O(n^2)$ steps with probability $1-O(1/n^3)$.
Therefore, at least one of the four segments disappears within $O(n^2)$ steps
with probability $1-O(1/n^3)$.
\end{proof}

By Claim~\ref{claim:disappear} and the union bound,
the number of segments decreases by a constant factor within every $O(n^2)$ steps with probability $1-O(1/n^2)$.
Hence, after $O(\log n)$ phases, only one segment remains, i.e., the ring orientation is achieved.
Therefore, $\por$ converges within $O(n^2 \log n)$ steps w.h.p.\ and thus in expectation.
\end{proof}

By Theorems~\ref{theorem:main} and~\ref{theorem:ssro},
we obtain the following undirected version of Theorem~\ref{theorem:main}.

\begin{theorem}
\label{theorem:undirected}
If an integer $\psi$ such that $\log n \le \psi\le \log n + O(1)$ is given,
there exists a self-stabilizing leader election protocol $P$ that works for any undirected ring such that (i) for any $C_0 \in \call(P)$,
$\Xi_{P}(C_0,\sch)$ reaches a safe configuration within $O(n^2 \log n)$ steps
both w.h.p.\ and in expectation, and (ii) each agent uses $\polylog(n)$ states.
\end{theorem}

\begin{proof}
We execute $\ppl$ and $\por$ in parallel.
Specifically, whenever two agents $u$ and $v$ interact as the initiator and the responder, respectively,
they first execute $\por$.
Thereafter, if $u.\dir = v.\iro$, they execute $\ppl$ as it is.
Otherwise, they execute $\ppl$ after swapping their roles as the initiator and the responder. 

By Theorem~\ref{theorem:ssro}, the ring orientation stabilizes within $O(n^2 \log n)$ steps. 
Then, by Theorem~\ref{theorem:main}, within the next $O(n^2 \log n)$ steps, 
the population reaches a safe configuration from which the same unique leader is preserved forever.
\end{proof}

\section*{Acknowledgements}
This work is supported by JST FOREST Program JPMJFR226U and JSPS KAKENHI Grant Numbers JP20KK0232, JP25K03078, JP25K03079, JP25K03101, and JP25K15057.

\bibliographystyle{alphaurl} 
\bibliography{biblio}

\appendix
\clearpage

\section{Chernoff bounds}
\label{sec:chernoff}
\begin{lemma}[\cite{kyoukasyo}, Theorems 4.4, 4.5]
\label{lem:chernoff}
Let $X_1,\dots,X_s$ be independent Poisson trials,
and let $X = \sum_{i=1}^s X_i$.
Then
\begin{align}
 \label{eq:upperdouble}
 \forall \delta,~0 \le \delta \le 1:~
 \Pr(X \ge (1+\delta)\ex[X]) &\le e^{-\delta^2\ex[X]/3},\\
 \label{eq:lowerhalf}
 \forall \delta,~0 < \delta < 1:~
 \Pr(X \le (1-\delta)\ex[X]) &\le e^{-\delta^2\ex[X]/2}.
\end{align}
\end{lemma}

\begin{algorithm}[t]
\caption{$\pcl$\ \ \ $a$ and $b$ are the initiator and responder of an interaction, respectively.}
\label{al:two-hop}
\VarAgent{
\\
\varspace $\iro, \scol \in [0,15]$, $\sbin \in \{0,1\}$
}
\If{$(a.\iro,a.\scol,a.\sbin)=(b.\scol,b.\iro,1-b.\sbin)$}{
$(a.\iro,b.\iro)
\gets
\text{two independent uniform random integers from } [0,15]$\;
\tcp*{Can be derandomized; see Note~\ref{note:derandomize}.}
}
$(a.\scol,b.\scol) \gets (b.\iro,a.\iro)$\;
$(a.\sbin,b.\sbin) \gets (1-a.\sbin,1-a.\sbin)$\;
\tcp*{Synchronize $\sbin$ while flipping the initiator's bit.}
\end{algorithm}

\section{Two-hop coloring}
\label{sec:two-hop}
In this section, we present a slight modification of the self-stabilizing
two-hop coloring protocol given by Sudo~\etal~\cite{SOK+18}.
This variant is designed specifically for rings and uses only $O(1)$ states.
We prove that it converges in $O(n^2)$ steps in expectation.
Although we conjecture that its actual convergence time is much shorter,
we do not pursue a tighter bound, since the $O(n^2)$ bound is sufficient
for the SS-RO protocol presented in Section~\ref{sec:orient}.
This protocol uses randomization, but can be derandomized without increasing
time or space complexities, as we will see in Note~\ref{note:derandomize}.

The pseudocode of the resulting protocol $\pcl$ is presented in
Algorithm~\ref{al:two-hop}.
Each agent $v$ maintains its current color $v.\iro \in [0,15]$, a color
stamp $v.\scol \in [0,15]$, and a binary stamp $v.\sbin \in \{0,1\}$.
Whenever two neighboring agents $u_i$ and $u_{i+1}$ interact, they store
each other's colors in their color stamps and synchronize their binary
stamps while flipping the initiator's binary stamp (Lines~79--80).
Thus, immediately after the interaction,
\[
(u_i.\iro,u_i.\scol,u_i.\sbin)
=
(u_{i+1}.\scol,u_{i+1}.\iro,u_{i+1}.\sbin).
\]
Before updating these stamps, $u_i$ and $u_{i+1}$ check whether
\[
(u_i.\iro,u_i.\scol,u_i.\sbin)
=
(u_{i+1}.\scol,u_{i+1}.\iro,1-u_{i+1}.\sbin).
\]
If there were no color conflict between $u_{i-1}$ and $u_{i+1}$ or
between $u_i$ and $u_{i+2}$, the first two equalities would imply that
neither $u_i$ nor $u_{i+1}$ has interacted with its other neighbor since
their last interaction.
Their binary stamps would therefore still be equal, contradicting the
third equality.
Hence, the above condition implies the existence of a color conflict at
distance two.
Then, $u_i$ and $u_{i+1}$ independently choose new colors uniformly at
random from $[0,15]$ (Lines~77--78).

A stamp conflict may be spurious, either because of an arbitrary initial
configuration or because the underlying color conflict has already been resolved
by recoloring.
Nevertheless, our analysis accommodates such spurious conflicts and treats
every stamp conflict as a defect to be eliminated.

\begin{note}[Derandomization]
\label{note:derandomize}
One can derandomize this process using the technique presented by
Burman~\etal~\cite{BCC+26}.
Each agent maintains a role in $\{\Alg,\Flip\}$ and switches its role
whenever it participates in an interaction.
When an agent needs a random bit, it waits for an interaction in which
it is in role $\Alg$ and its interaction partner is in role $\Flip$.
The bit is set to $0$ if the agent is the initiator and to $1$ if it is
the responder.
As observed in~\cite{BCC+26}, this yields a fair random bit independently
of the state of the interaction partner and of the previously generated
random bits.
Repeating this procedure four times yields a uniform random integer in
$[0,15]$.
Our protocol uses randomness only when two agents $a$ and $b$ detect a
stamp conflict.
Before reusing randomness, each agent must generate four fresh random bits.
Thus, if either $a$ or $b$ has not yet generated four fresh bits when a
stamp conflict is detected, the recoloring operation in Line~78 is simply
skipped, while the remaining commands are executed as usual.
Since generating four fresh random bits requires only $O(n)$ steps in
expectation, this modification does not affect the asymptotic bounds in
the subsequent analysis.
\end{note}

\begin{definition}
Two agents $u_i$ and $u_{i+2}$ have a \emph{color conflict} if
$u_i.\iro = u_{i+2}.\iro$.
\end{definition}

\begin{definition}
Two neighboring agents $u_i$ and $u_{i+1}$ have a \emph{stamp conflict} if
\[
(u_i.\iro,u_i.\scol,u_i.\sbin)
=
(u_{i+1}.\scol,u_{i+1}.\iro,1-u_{i+1}.\sbin).
\]
\end{definition}

\begin{definition}
An agent $u_i$ is \emph{red} if
\begin{itemize}
\item $u_i$ has a color conflict with $u_{i-2}$ or $u_{i+2}$, 
\item $u_i$ has a stamp conflict with $u_{i-1}$ or $u_{i+1}$, or
\item $u_{i-1}$ and $u_{i+1}$ have a color conflict.
\end{itemize}
Otherwise, $u_i$ is \emph{blue}.
\end{definition}

\begin{lemma}
\label{lem:red_increase}
At each step, a blue agent becomes red only when a stamp conflict is detected.
\end{lemma}

\begin{proof}
Consider an interaction between $u_i$ and $u_{i+1}$ at which they do not have a stamp conflict.
Then, only their stamps are updated, so no new color conflict is
created.
Moreover, $u_i$ and $u_{i+1}$ do not have a stamp conflict after the
interaction because their binary stamps are synchronized.
Hence, a new stamp conflict can arise only between $u_{i-1}$ and $u_i$
or between $u_{i+1}$ and $u_{i+2}$.
In the former case, $u_{i-1}.\iro=u_{i+1}.\iro$, so
$u_{i-1}$ and $u_{i+1}$ already have a color conflict; hence,
$u_{i-1}$ and $u_i$ are already red.
The latter case is symmetric.
Thus, no blue agent becomes red.
\end{proof}

\begin{lemma}
\label{lem:red_decrease}
If there is at least one red agent, one of the following events occurs
within $O(n)$ steps in expectation:
\begin{itemize}
\item two agents detect a stamp conflict and recolor themselves, or
\item the number of red agents decreases by at least one.
\end{itemize}
\end{lemma}

\begin{proof}
Suppose that there is at least one red agent.
By Lemma~\ref{lem:red_increase}, it suffices to show that, within $O(n)$
steps in expectation, either a stamp conflict is detected or a red agent
becomes blue.

Suppose first that there is a color conflict,
say between $u_{i-1}$ and $u_{i+1}$.
Let $e_1,e_2,e_3$ be
the next three interactions involving at least one of
$u_{i-1},u_i$, and $u_{i+1}$.
The event
\[
e_1= (u_{i-1},u_i),\qquad
e_2=(u_i,u_{i+1}),\qquad
e_3=(u_{i-1},u_i)
\]
occurs with probability $(1/8)^3=\Theta(1)$.
Since an interaction involving at least one of these three agents occurs within
$O(n)$ steps in expectation, this event occurs within $O(n)$ steps in
expectation.
During these three interactions, $u_i$ detects a stamp conflict,
unless $u_{i-1}$ or $u_{i+1}$ detects a stamp conflict with another agent
earlier.
Hence, a stamp conflict is detected within $O(n)$ steps in expectation.

Suppose next that there is no color conflict.
Since there is a red agent, there must be a stamp conflict, say between
$u_i$ and $u_{i+1}$.
Within $O(n)$ steps in expectation, $u_i$ interacts with either
$u_{i-1}$ or $u_{i+1}$.
In the former case, since there is no color conflict, no new stamp conflict
is created, and hence $u_i$ becomes blue.
In the latter case, $u_i$ detects the stamp conflict.
Thus, in either case, one of the desired events occurs within $O(n)$ steps
in expectation.
\end{proof}

\begin{theorem}
\label{theorem:two-hop}
Protocol $\pcl$ is a self-stabilizing two-hop coloring protocol using
$O(1)$ states.
Starting from any configuration, an execution of $\pcl$ reaches,
within $O(n^2)$ steps in expectation, a configuration in which
$u_i.\iro \neq u_{i+2}.\iro$ for every $i \in [0,n-1]$ and from which
no agent ever changes its $\iro$ variable.
\end{theorem}

\begin{proof}
We say that a configuration is \emph{red-free} if no agent is red.
In a red-free configuration, there is no stamp conflict, so no agent can
become red by Lemma~\ref{lem:red_increase}.
Hence, the set of red-free configurations is closed.
It therefore suffices to bound the expected time until the population
reaches a red-free configuration.

We consider the following two types of events:
\begin{itemize}
\item a \emph{type-I event}, in which a stamp conflict is detected and
the two interacting agents recolor themselves, and
\item a \emph{type-II event}, in which no stamp conflict is detected
but the number of red agents decreases by at least one.
\end{itemize}
We collectively call them \emph{good events}.

Consider a type-I event involving $u_i$ and $u_{i+1}$.
Both agents are red immediately before the event because they have a
stamp conflict.
There are at most four possible new color conflicts caused by their new
colors, and each corresponding color equality occurs with probability
$1/16$.
Thus, by the union bound, with probability at least $1-4/16=3/4$,
no such color conflict is created.
In this case, both $u_i$ and $u_{i+1}$ become blue, and no previously
blue agent becomes red.
Hence, the number of red agents decreases by at least two.
Otherwise, at most the four agents
$u_{i-2},u_{i-1},u_{i+2},u_{i+3}$ can change from blue to red.
Therefore, conditioned on the configuration immediately before any
type-I event, the expected decrease in the number of red agents is at
least
\[
\frac34\cdot 2-\frac14\cdot 4=\frac12.
\]
At every type-II event, the number of red agents decreases by at least
one.

Let $T$ be the number of good events until the population becomes
red-free.
For each $j\ge1$ such that $T\ge j$, let $C_j$ be the configuration
immediately before the $j$-th good event, and let $D_j$ be the decrease
in the number of red agents caused by this event.
For $T<j$, define $D_j=0$.
As argued above, for every possible $C_j$,
$\ex[D_j\mid C_j,T\ge j]\ge1/2$.
Hence,
\[\ex[\mathbf{1}_{T\ge j}D_j]
\ge \frac{1}{2} \ex[\mathbf{1}_{T \ge j}]
= \frac{1}{2}\Pr(T\ge j).\]
Therefore, for every $m\ge1$,
\[
n\ge\sum_{j=1}^{m}\ex[\mathbf{1}_{T\ge j}D_j]
\ge \frac{1}{2}\sum_{j=1}^{m}\Pr(T\ge j).
\]
Letting $m\to\infty$ yields $\ex[T]\le2n$.
Thus, the expected number of good events until a red-free configuration
is reached is $O(n)$.

Finally, by Lemma~\ref{lem:red_decrease}, whenever at least one red
agent exists, the expected number of steps until the next good event is
$O(n)$. Hence, the expected number of steps until a red-free configuration is
reached is $O(n)\cdot\ex[T]=O(n^2)$.
\end{proof}

\end{document}